\documentclass[sigconf,nonacm]{acmart}

\usepackage{array}
\usepackage{calc}
\usepackage{listings}
\usepackage[inline]{enumitem}
\usepackage{tabularx}
\usepackage{xcolor}
\usepackage{subcaption}

\usepackage{pifont}
\usepackage{xcolor}
\newcommand{\cmark}{\textcolor{green!70!black}{\ding{51}}}
\newcommand{\xmark}{\textcolor{red}{\ding{55}}}

\newcolumntype{L}[1]{>{\raggedright\arraybackslash}p{#1}}
\newcolumntype{C}[1]{>{\centering\arraybackslash}p{#1}}
\newcolumntype{Y}{>{\raggedright\arraybackslash}X}
\makeatletter
\let\ACM@orig@startsection\@startsection
\def\@startsection#1#2#3#4#5#6{%
  \ifstrequal{#1}{section}{%
    \ACM@orig@startsection{#1}{#2}{#3}%
      {-.55\baselineskip \@plus -1.5\p@ \@minus -.1\p@}%
      {.18\baselineskip}{#6}%
  }{%
    \ifstrequal{#1}{subsection}{%
      \ACM@orig@startsection{#1}{#2}{#3}%
        {-.55\baselineskip \@plus -1.5\p@ \@minus -.1\p@}%
        {.18\baselineskip}{#6}%
    }{%
      \ifstrequal{#1}{subsubsection}{%
        \ACM@orig@startsection{#1}{#2}{#3}%
          {-.4\baselineskip \@plus -1.2\p@ \@minus -.1\p@}%
          {-2.5\p@}{#6}%
      }{%
        \ACM@orig@startsection{#1}{#2}{#3}{#4}{#5}{#6}%
      }%
    }%
  }%
}
\makeatother

\newcommand{\husdraft}[1]{\textcolor{black}{#1}}

\newcommand{\rev}[1]{\textcolor{black}{#1}}

\AtBeginDocument{%
  }

\setcopyright{none} 
\copyrightyear{2026}
\acmYear{2026}
\acmDOI{XXXXXXX.XXXXXXX}

\acmConference[MICRO 2026]{The 59th IEEE/ACM International Symposium on Microarchitecture}{October 31--November 04, 2026}{Athens, Greece}

\acmISBN{978-X-XXXX-XXXX-X/XX/XX}

\begin{document}

\title{From Bit-Position Sensitivity to Unequal Error Protection for DNN Inference Memory}

\author{Muhammad Husnain Mubarik, Karthik Mohan Kumar, Pedro Antonio Pena, Keshavan Varadarajan, Kunal Tyagi}
\affiliation[obeypunctuation=true]{%
  \institution{Advanced Micro Devices, Inc.,}
  \country{USA}
}




\begin{abstract}
We characterize per-bit-position fault sensitivity in ML inference across 16
workloads --- spanning transformer-based models and attention-free CNNs ---
and across three floating-point formats. \rev{Our central empirical finding is
a sharp bit-sensitivity transition: flipping any of the least-significant fraction
 bits up to a data-type-specific threshold, \texttt{Xsafe}, degrades task metrics 
 by less than 1\% under deterministic single-bit stress tests. Sensitivity rises 
 through the upper fraction bits and spikes at the exponent-mantissa boundary, 
 where a single-bit flip causes catastrophic collapse. Because low-order bits are
 largely inconsequential while high-order and exponent bits are critical, uniform 
 SECDED protection---which guards every bit equally at 12.5\% storage 
 overhead---is unnecessarily conservative.}
\rev{We derive per-data-type \texttt{Xsafe} floors \{FP16:~6, BF16:~4,
FP32:~15\} and workload-aware tiers that widen the unprotected region for
resilient model classes, raising ECC savings to 37.5--62.5\% without
retraining. Text-conditioned diffusion models dictate the conservative floor;
vision encoders, NLU models, and resilient LLMs tolerate wider bypass regions.}
These floors and tiers drive an Unequal Error Protection (UEP) codec with per-cacheline data-type tags and
a dual-partition SRAM architecture for ML accelerators. Validation across 870+
fault-injection runs confirms selective protection holds under contiguous 2-
and 3-bit upsets. \rev{The codec reduces ECC area by 27.8\% relative to uniform
SECDED; dual-voltage operation of the non-critical partition lowers gross
BF16 read energy by ${\sim}17\%$, with a ${\sim}4\%$ dual-partition
macro-area overhead.}
\end{abstract}



\maketitle

\renewcommand{\thefootnote}{}%
\footnotetext{This is the authors' preprint version of a paper accepted to appear at the 59th IEEE/ACM International Symposium on Microarchitecture (MICRO 2026).}%
\renewcommand{\thefootnote}{\arabic{footnote}}%

\section{Introduction}

Deep Neural Network (DNN) inference is increasingly deployed in environments with  
frequent memory faults: radiation-induced soft errors in high-reliability 
accelerators~\cite{sullivanGpuDram2021,sridharan2015memory}, retention failures in 
scaled main-memory technologies~\cite{qureshi2015,mathew2018,oh2016,schroeder2009dramwild,meza2015revisiting},
and bit-flips from low-voltage SRAM operations~\cite{stutz2021}. 
The standard response is uniform error correction, protecting every stored bit 
equally. Conventional Hamming SECDED adds 8 check bits per 64 data bits
(12.5\% storage overhead)~\cite{hamming1950,hsiao1970,tang2016,mathew2018}.
With the ever-increasing compute capacity and consequent large power 
dissipation in current ML hardware, the compute and area footprint required for this uniform protection becomes an 
increasingly expensive and (potentially unnecessary) guarantee for every bit.

Since IEEE 754~\cite{ieee754_2019,goldberg1991} encodes magnitude non-uniformly, this asymmetry raises three questions. 
\begin{enumerate*}
    \item Can we protect only some bits (e.g., the sign and exponent bits) and leave other
    bits unprotected?
    \item Is there a threshold number of unprotected bits below 
    which task-quality metrics remain acceptable?
    \item Is that threshold universal across model architectures and data types -- 
    and therefore usable as a hardware design parameter?
\end{enumerate*}

In this paper, we present a comprehensive per-bit-position fault-sensitivity study across 16 models, 6 modalities, and three floating-point formats (FP16, BF16, FP32) to derive conservative per-data type floor for bits that can safely remain unprotected (\S~\ref{fault-modeling-and-experimental-setup}). We propose workload-aware tiers that trade additional mantissa-bit exposure for greater ECC savings (\S~\ref{bit-position-sensitivity-results}). The existence of these thresholds motivates Unequal Error Protection (UEP)—an error-correction strategy that assigns different protection levels to different bits in the data word \cite{UEPBook}. We design a UEP codec and per-cacheline tag architecture (\S~\ref{sec:architecture}) based on these thresholds and validate the design under multi-bit fault patterns and evaluate SRAM-level area, energy, and reliability impact
(\S~\ref{design-space-exploration-and-reliability}). \S~\ref{related-work} surveys prior approaches and shows that this is the first work 
to identify bit-protection floors across data types and workloads and co-design them
into a UEP codec with SRAM-level partitioning. Our contributions are: 
\begin{enumerate}[leftmargin=*, topsep=0pt]
  \item \emph{Per-bit-position fault-sensitivity characterization}: We systematically study the effect of single-bit faults at every bit position, identify conservative per-data-type thresholds for the number of fraction LSBs that can safely remain unprotected, and define workload-aware tiers that further relax protection for more resilient model classes.
  \item \emph{UEP codec and tag architecture}: We design a UEP codec that assigns graded protection by bit significance (e.g., full SECDED on sign and exponent bits, single-error correction on high-order fraction bits, and no protection on tolerant fraction LSBs) controlled by per-cacheline data-type tags.
   \item \emph{Multi-bit fault validation}: We validate the selective protection scheme under contiguous 2- and 3-bit upsets across 870+ fault-injection experiments and show that bit-plane interleaving eliminates boundary-crossing failures between protected and unprotected bit regions.
   \item \emph{SRAM-level hardware evaluation.} \rev{We synthesize the codec in RTL and model the SRAM macro to evaluate area, read energy, and reliability. The UEP codec reduces ECC codec area by 27.8\% relative to uniform SECDED; dual-voltage operation of the non-critical partition reduces BF16 read energy by \({\sim}17\%\) (gross; net positive after level shifters, no precision loss), while the dual-partition macro adds \({\sim}4\%\) area that is recovered by these energy savings,} \husdraft{with all SER-regime configurations remaining below 10 failures in time (FIT) and the dual-voltage operating point maintaining an MTTF of \(\sim\!\)900 years}.
\end{enumerate}
\section{Related Work}\label{related-work}

\subsection{DNN fault tolerance and approximate
computing}\label{dnn-fault-tolerance-and-approximate-computing}

Prior work establishes that DNN fault tolerance depends strongly on
where corruption lands, but it stops short of the per-bit-position,
cross-data type characterization developed in this paper.

Ares~\cite{reagen2018} and ThUnderVolt~\cite{zhang2018} (34--57\% energy savings
at \textless1\% accuracy loss) quantify DNN
vulnerability per layer and accelerator structure yet treat all bits within
a floating-point word uniformly.

Approximate-memory based approaches move closer to value
awareness: EDEN~\cite{koppula2019}, the Approximate Memory
architecture~\cite{approxMem2018}, Stutz et al.~\cite{stutz2021},
Catalán et al.~\cite{catalan2025}, and FIGARO~\cite{wang2020} achieve
~20--37\% energy savings at \textless1\% accuracy loss, but characterize
tolerance at word, column, cache-block, or weight level rather than per
bit position within a floating-point word, and most require retraining.
\husdraft{Security-oriented fault-attack and accelerator-resilience studies reach a similar conclusion from another angle: targeted single-bit faults in model parameters can be catastrophic~\cite{hong2019terminal}, and propagation through DNN accelerators depends strongly on data type, value range, reuse, and layer type~\cite{li2017errorprop}.}

This paper differs from all of the above in three ways: (1)
per-bit-position characterization,
(2) no retraining, and (3) actionable thresholds that translate
directly to ECC design parameters.

\subsection{Selective and adaptive ECC for
memory}\label{selective-and-adaptive-ecc-for-memory}

Tang et al.~\cite{tang2016} combine repetition codes with Bose--Chaudhuri--Hocquenghem (BCH) codes for unequal error
protection of embedded SRAM in DSPs, assigning stronger ECC to
high-order bits (HOB) and lighter ECC to low-order bits (LOB). This is
direct prior art for bit-position-selective ECC. However, the HOB/LOB
split is arbitrary (not derived from application-level sensitivity
data), the scheme requires two codec passes, and the tier boundaries are
not optimized for floating-point bit-field structure.

SERA-Float~\cite{mishra2025}
protects sign+exponent with ECC and stores 8 fraction bits with lossy-compression,
achieving up to 80.3\% energy savings per multiply at 0.9\% accuracy
loss. It is focused on FP32 and uses three independent codecs for one value while our work achieves similar guarantees with a single lossless codec.

\rev{A single fixed, data type-agnostic protection boundary cannot be both safe
and efficient across datatypes. SERA-Float commits one \(X{=}8\) (FP32-tuned)
policy; Table~\ref{tab:related-sera-x-comparison} contrasts it with our
empirically derived per-data type \texttt{Xsafe} floor. Because \(X{=}8\) exceeds
the FP16 and BF16 floors, it leaves task-critical bits unprotected for the
most sensitive models, yet it still over-protects FP32 where 15 low bits are
provably safe to bypass.}

\begin{table}[htbp]
\centering
\footnotesize
\setlength{\tabcolsep}{4pt}
\renewcommand{\arraystretch}{1.0}
\caption{\rev{SERA-Float's fixed \(X{=}8\) policy versus our per-data type \texttt{Xsafe} floor.}}
\label{tab:related-sera-x-comparison}
\begin{tabularx}{\linewidth}{@{}C{0.10\linewidth} C{0.10\linewidth} C{0.16\linewidth} Y@{}}
\toprule
data type & SERA \(X\) & Our floor \texttt{Xsafe} & Fixed-policy error of SERA \(X{=}8\) \\
\midrule
FP16 & 8 & 6  & Exposes critical bits in low-\texttt{Xsafe} diffusion \\
BF16 & 8 & 4  & Unsafe: \(X{=}8\) exceeds BF16's 7 mantissa bits \\
FP32 & 8 & 15 & Over-protects: 7 safe bits wasted \\
\bottomrule
\end{tabularx}

\end{table}

Several adaptive ECC schemes adjust protection strength without
data type awareness. Stealth ECC~\cite{lee2022} varies correctability by
data width (47.9\% failure-probability reduction, 0.9\% performance
overhead); COMET~\cite{alam2022} co-designs on-die SEC with
in-controller SECDED; AFT-ECC~\cite{aftEcc2023} repurposes check bits
for metadata; Zheng et al.~\cite{zheng2017} protect memory subsets at
page-level granularity. None distinguishes floating-point bit positions.

GPU-centric precedents support structure-local deployment:
Palframan et al.~\cite{palframan2014} explore precision-aware
soft-error protection for GPU register files, and
AFT-ECC~\cite{aftEcc2023} shows that ECC-path bits can carry metadata
without a separate side array. Our tiers derive from per-bit-position
DNN sensitivity rather than static precision classes, but the mechanisms
are compatible.

At the rank and controller level, Chipkill RS~\cite{ha2025} uses (10,8)
4-bit symbol Reed--Solomon codes for DDR5 chipkill-level protection;
Frugal ECC~\cite{frugalEcc} frees space for stronger check bits via
fine-grained compression; Breaking the HBM Bit Cost
Barrier~\cite{xie2025bitcost} moves Reed--Solomon protection into the
memory controller with an importance-adaptive bit-plane policy; and
REACH~\cite{xie2025} extends this with two-level erasure coding. All
four use Reed--Solomon or system-level codes for DRAM/HBM and are
orthogonal to our work: they target
raw channel bit-error rate (BER) at the rank or controller level (and can leverage our per-cacheline format tag), whereas UEP derives
per-bit-position protection tiers from measured task-level sensitivity.

Across these prior schemes, the key distinction is therefore
the protection boundary. The construction here uses a single-codeword,
weight-partitioned parity-check matrix whose tier boundaries come from
per-bit-position DNN sensitivity data across multiple data types rather
than from fixed width classes, page categories, or controller-scale HBM
codeword spans.
\rev{The distinguishing novelty is thus threefold: the tier boundaries are
derived empirically from per-bit-position task sensitivity (16 models,
6 modalities, 3 data types) rather than from a heuristic split or a single fixed
policy; this is the first precision-aware selective-ECC scheme co-designed for
SRAM bit positions; and the resulting floor is validated end-to-end on task
metrics across modalities and under multi-bit upsets, beyond the single-bit
``exponent matters'' observation.}

\vspace{-0.1in}
\section{Fault modeling and experimental setup}\label{fault-modeling-and-experimental-setup}

\subsection{Fault model}\label{fault-model}
Memory faults in this study fall into two domains: continuous analog noise and
discrete digital bit-flips.

\textbf{Continuous noise} is parameterized along two axes:
\begin{enumerate}[topsep=0pt, leftmargin=*]
    \item \emph{Coupling mode:} relative noise scales with the signal
    ($\tilde{x}=x(1+\epsilon)$), while constant noise is signal-independent
    ($\tilde{x}=x+\epsilon$). These map to effects such as voltage-dependent SRAM
    errors and quantization rounding, respectively.
    \item \emph{Distribution shape:} Gaussian (thermal), uniform
    (quantization), log-uniform (continuous relaxation of mantissa flips),
    power-law $1/f$ (flicker~\cite{joshi2020pcm}), and lognormal
    (device variation~\cite{narayanan2017nvm,joshi2020pcm}).
\end{enumerate}

\textbf{Discrete faults} inject bit-flips under three complementary protocols:
\begin{enumerate}[topsep=0pt, leftmargin=*]
    \item \emph{Deterministic fixed-position stress tests:} one fixed bit is flipped
    in every floating-point element to isolate per-position sensitivity.
    \item \emph{Stochastic random-position sweeps:} one random bit/word is flipped
    with probability $p_{word}$, approximating realistic fault rates.
    \item \emph{Correlated multi-bit patterns:} contiguous 2- or 3-bit flips model
    multi-cell upsets from one strike.
\end{enumerate}
Transient faults (single-bit SEUs~\cite{sullivanGpuDram2021,sridharan2015memory})
isolate per-layer vulnerability; persistent faults (stuck-at behavior from
aging/retention effects~\cite{qureshi2015,mathew2018}) corrupt layers \(P\)
through \(P+i\). Continuous campaigns sweep coupling/distribution effects on
task metrics, while discrete campaigns locate the benign-to-catastrophic cliff location
and test whether stochastic, correlated, or persistent faults shift it.

\textbf{Primary fault combinations:} Each campaign pairs a physically 
motivated coupling mode with a distribution or flip pattern in Table 
\ref{tab:campaign_distribution}.
\begin{table}[h]
    \centering
    \footnotesize
    \setlength{\tabcolsep}{4pt}
    \newcolumntype{P}[1]{>{\raggedright\arraybackslash}p{#1\columnwidth}}
\caption{Primary fault combinations tested. Each campaign pairs a coupling mode with a distribution or flip pattern, mapped to its physical analog.}
\label{tab:campaign_distribution}
    \begin{tabular}{P{0.27}P{0.28}P{0.4}}\toprule
      \textbf{Campaign} &  \textbf{Fault Mode} & \textbf{Physical analog} \\\midrule
      SRAM Voltage Scaling & Relative $\times$ Gaussian &  Voltage-dependent (BER) \cite{stutz2021} \\
      Fraction relaxation & Relative $\times$ log-uniform & Continuous approximation of random fraction flips \\
      Quantization rounding & Constant $\times$ uniform & Narrow-format arithmetic \\
      Cell variation & Constant $\times$ lognormal & Device-to-device SRAM mismatch \\
      Radiation SEU & Discrete random $\times$ transient & Single-event upsets \\
      Aging / retention & Discrete fixed-value $\times$ persistent & Stuck-at faults \\
    \bottomrule\end{tabular}
    
\end{table}

Two additional matched-variance shape-analysis cases (relative 
log-normal and pink ($1/f$)) are included in \S~\ref{sec:continuous-results} to test if 
degradation depends on distribution shape independently of variance.

\subsection{Models and metrics}\label{models-and-metrics}
The evaluation spans \husdraft{16 models across 6 modalities}. \husdraft{The
set is chosen to test both cross-modal generality and
attention-independence: attention-free CNNs serve as controls for whether the
cliff follows attention structure or floating-point data type.
Table~\ref{tab:modalities-models-metrics} lists modalities, models, attention
types, and primary metrics; per-model \texttt{Xsafe} outcomes are reported in
Table~\ref{tab:cross-arch-xsafe}.}

\begin{table}[htbp]
\centering
\footnotesize
\setlength{\tabcolsep}{4pt}
\renewcommand{\arraystretch}{1.0}
\caption{\husdraft{Evaluation modalities, models, attention mechanisms, and
primary task metrics. Attention abbreviations: MHA (multi-head attention),
MMDiT (multimodal diffusion transformer), GQA (grouped-query attention), SWA
(sliding-window attention), MQA (multi-query attention), and SE
(squeeze-and-excitation).}}
\label{tab:modalities-models-metrics}
\begin{tabularx}{\linewidth}{@{}L{0.14\linewidth} Y L{0.16\linewidth} Y@{}}
\toprule
\textbf{Modality} & \textbf{Models} & \textbf{Attention types} & \textbf{Metric} \\
\midrule
Image generation & SD3.5-Medium (2.5B), DiT-XL/2 (675M), SD 1.5 (860M),
PixArt-$\alpha$ (611M) & MHA, MMDiT, MHA+Cross & PSNR / SSIM~\cite{wang2004ssim} / LPIPS \\
\husdraft{Text generation} & \husdraft{Llama-3 8B, Phi-3-mini (3.8B), Mistral-7B (7.3B),
GPT-2 Medium (355M), Falcon-7B (7B)} & \husdraft{GQA, GQA+SWA, MHA, MQA}
& \husdraft{WikiText-2 perplexity (2048-token, headline 10-segment 95\% confidence interval (CI))} \\
Vision classification & ViT-L/16 (304M), Swin-B (88M) &
MHA, shifted window & ImageNet-1K top-1 accuracy (5,000
images) \\
Language understanding & BERT-base (110M) & MHA (bidirectional) & SST-2
classification accuracy (872 examples) \\
Speech recognition & Whisper-small (244M) & MHA+Cross & LibriSpeech
test-clean WER (200 utterances) \\
CNN image classification & ResNet-50 (25M), EfficientNet-B0 (5.3M),
MobileNetV3-Large (5.4M) & None, SE (channel) & ImageNet-1K top-1
accuracy (5,000 images) \\
\bottomrule
\end{tabularx}

\end{table}

\husdraft{Model/architecture provenance and decoder-attention labels
(MQA/GQA/SWA) follow the original publications~\cite{rombach2022ldm,peebles2023dit,sdxl2023,esser2024rf,chen2023pixartalpha,playground2024,sauer2024ladd,radford2019gpt2,falcon2023,mistral7b_2023,phi3_2024,llama3_2024,devlin2019bert,radford2022whisper,dosovitskiy2021vit,liu2021swin,he2016resnet,tan2019efficientnet,howard2019mobilenetv3,mqa2019,gqa2023}. Dataset provenance follows original releases~\cite{russakovsky2015imagenet,socher2013sst,panayotov2015librispeech,merity2017pointer}; C4~\cite{dodge2021c4} and HellaSwag~\cite{zellers2019hellaswag} provide external-validity checks. Auxiliary diffusion checkpoints used in focused analyses do not change the cross-model floors in Table~\ref{tab:cross-arch-xsafe}.}

For generative models, LPIPS~\cite{zhang2018lpips} is the preferred perceptual
distance metric. KL divergence, rank correlation, and top-k stability
complement primary metrics, and robustness curves plot each metric
against perturbation severity. \husdraft{When later sections call a bit
position ``safe'' or report an \texttt{Xsafe} floor, the criterion is always
evaluated relative to that modality's clean baseline metric in
Table~\ref{tab:modalities-models-metrics}; we do not compare raw percentage
changes across disparate metrics.}

\husdraft{For synthetic layer-wise profiling of vision transformers and CNNs,
top-1 accuracy is not meaningful, so we report scaled logit cosine divergence:
\(D_{\mathrm{logit}} = 100\,(1 - \cos(z_{\mathrm{clean}}, z_{\mathrm{faulty}}))\),
averaged across input logits ($z$). Since cosine similarity lies in \([-1, 1]\),
\(D_{\mathrm{logit}}\in[0,200]\), values above 100 indicate mean negative
cosine similarity.}

\husdraft{For the main decoder-LLM case study, we also run a compact
external-validity check on C4 and HellaSwag with the same Llama-3~8B BF16
attention injector, so conclusions are not anchored to WikiText-2 alone.}

\subsection{Injection protocol}\label{injection-protocol}
\textbf{Design principle.} Every campaign keeps \emph{fault family} and \emph{injection site}
conceptually separate: the fault family (continuous vs.\ discrete, transient 
vs.\ persistent) defines \emph{what} is injected; the model family defines 
\emph{where}---i.e.\ which logical tensor is exposed.

\textbf{Injection sites.} For attention-based models the primary target 
is the pre-Softmax $QK^{\!\top}$ score tensor; when a fused-attention kernel 
hides the intermediate logits we materialize an explicit
$QK^{\!\top}\!\to\!\text{Softmax}\!\to\!V$ path. For non-attention paths and 
mixed architectures, we target the native Feed-Forward Network (FFN) / 
Multi-Layer Perceptron (MLP) block outputs (decoder-LLM MLP blocks, ViT/Swin 
feed-forward blocks, BERT intermediate layers, Whisper FFN, diffusion FFN modules).
Convolutional Neural Networks (CNN) are injected at convolutional blocks and 
the final classifier. Additional loci---weights, activations, inputs, and 
normalization layers---are tested for selected models only when a within-model
sensitivity ranking is needed.

\textbf{Fault campaigns.} Each injection site is tested against every fault
family in \S~\ref{fault-model} and Table~\ref{tab:campaign_distribution}.
Continuous campaigns sweep noise amplitude (e.g.,
$\sigma\in\{1\%,5\%,10\%\}$) across coupling modes/distributions. Discrete
campaigns cover transient and persistent axes in single-event mode (inject at
layer~$P$, measure at $P{+}i$) and continuous-event mode (corrupt layers $P$
through $P{+}i$).

\textbf{Precision scope.} Campaigns run in FP16, BF16, and FP32 (native and
accumulation data types). FP8 formats are storage-only (accumulation still uses
FP16/BF16/FP32), so selective ECC targets long-lived FP16/BF16/FP32 values in
registers and SRAM. \husdraft{\texttt{Xsafe} denotes the number of fraction
LSBs that remain unprotected while staying within a modality-specific quality
bound: perplexity~${\leq}\,1.01{\times}$ baseline (LLMs),
accuracy~${\geq}$ baseline~$-\,1$\,pp (vision/NLU),
WER~${\leq}$ baseline~$+\,1$\,pp (speech), or the first bit before a
single-step PSNR drop \(>5\)\,dB (diffusion).} \texttt{Xsafe} is derived in
\S~\ref{bit-position-sensitivity-results} and characterized under both
full-precision and FP8-quantized workloads.

\subsection{Statistical protocol}\label{statistical-protocol}
Unless noted otherwise, each robustness condition uses at least 35 matched
inputs and is reported with 95\% confidence intervals; headline counts are
listed in Table~\ref{tab:modalities-models-metrics}. Key diffusion
\(p_{\text{word}}\) sweeps (\S~\ref{bit-position-sensitivity-results}) are
rerun with 100 prompts to validate the 35-sample choice.
Noise-shape comparisons (\S\ref{sec:continuous-results}) use paired Wilcoxon
signed-rank tests~\cite{wilcoxon1945}. Classification intervals (BERT, ViT)
use exact Clopper--Pearson binomial intervals~\cite{clopper1934}. Paired
significance cross-checks in \S~\ref{sec:datatype-property} use McNemar
tests~\cite{mcnemar1947}, one per bit position with Bonferroni correction
(family size = data type width: 16 for FP16/BF16, 32 for FP32).\footnote{To match
ViT's 5{,}000-image subset, the BERT McNemar study uses 5{,}000 SST-2 examples
(872 validation plus 4{,}128 training examples reserved for this diagnostic).}

\section{Bit-Position Sensitivity Results}\label{bit-position-sensitivity-results}
{
This section characterizes per-bit-position fault sensitivity under discrete
(\S~\ref{sec:discrete-results}) and continuous (\S~\ref{sec:continuous-results})
faults.

\subsection{Discrete Bit-Flip Sensitivity}\label{sec:discrete-results}

The subsections below systematically establish which bits can safely remain
unprotected, then stress-test that conclusion by relaxing one experimental
assumption at a time. \S~\ref{sec:xsafe-threshold} establishes the baseline
answer under controlled conditions: deterministic single-bit-position sweeps
at maximum stress isolate the contribution of each bit position individually
and yield a conservative per-data-type protection threshold, \texttt{Xsafe}.
\S~\ref{sec:stuck-at} relaxes the
transient-fault assumption, showing that \texttt{Xsafe} holds under permanent
stuck-at faults and characterizing the polarity asymmetry they introduce.
\S~\ref{sec:ber-sweeps} relaxes the deterministic and single-bit assumptions,
showing that random bit-flips at realistic fault rates respect the same
threshold. \S~\ref{sec:tiers} shows that the conservative floor can be widened
for more resilient workloads, yielding per-modality tiers that increase ECC
savings without retraining. \S~\ref{sec:mcu} relaxes the single-bit-per-event
assumption, showing that multi-cell upsets contract \texttt{Xsafe} by at most
one bit per additional flipped bit.

\begin{figure}[t]
\centering
\includegraphics[width=0.75\linewidth]{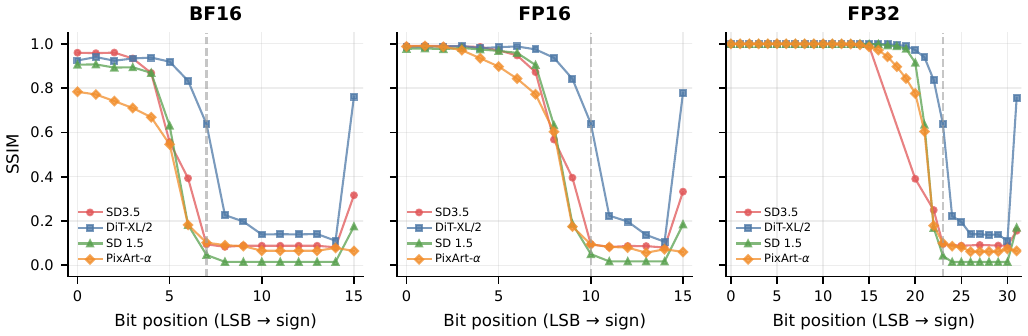}
\caption{Per-model SSIM under deterministic single-bit-position stress tests for
SD3.5-Medium, DiT-XL/2, SD~1.5, and PixArt-$\alpha$ across BF16 (left),
FP16 (center), and FP32 (right). Each point stresses one bit position across
all targeted floating-point words ($p_\text{word}=1$); dashed lines mark the
fraction/exponent boundary for each data type. Fraction LSBs up to
\texttt{Xsafe} cause negligible quality loss; sensitivity ramps through the
upper fraction bits before collapsing at the exponent field.
The data-type-specific onset of this ramp --- not the exponent boundary itself
--- defines \texttt{Xsafe}.}
\label{fig:cross-dtype-transition}
\end{figure}

\subsubsection{Per-Bit Sensitivity and the \texttt{Xsafe} Protection Threshold}\label{sec:xsafe-threshold}
\paragraph{Evidence from deterministic sweeps.}
Deterministic single-bit-position sweeps -- where each bit position is
individually stressed at a per-word fault rate $p_\text{word} = 1$ (every
floating-point word in the targeted tensor has that specific bit toggled) --
reveal a consistent two-phase sensitivity pattern.
Figure~\ref{fig:cross-dtype-transition} shows this pattern for four diffusion
models (SD3.5-Medium, DiT-XL/2, SD~1.5, and PixArt-$\alpha$) across BF16,
FP16, and FP32: corrupting the least-significant fraction bits causes negligible
quality loss, while sensitivity ramps through the upper fraction bits before
collapsing catastrophically once the exponent field is reached. The same
two-phase pattern holds across architecturally distinct model types, as shown
in Table~\ref{tab:margin-to-failure}: for Llama-3~8B, perplexity remains stable
through bit~11 but jumps from 7.01 to 98.6 at bit~12; for ViT-L/16, top-1
accuracy drops 3.94 percentage points over the same two-bit window; for
SD3.5-Medium, SSIM falls by 0.19 per step once the transition begins.

\paragraph{The \texttt{Xsafe} threshold.}
The onset of this sensitivity ramp defines a data-type-specific protection
threshold, \texttt{Xsafe}: the number of fraction LSBs that can safely remain
unprotected without crossing each modality's quality bound
(\S~\ref{injection-protocol}). Table~\ref{tab:cross-arch-xsafe} consolidates
\texttt{Xsafe} across all 16 evaluated models. The cross-model floor --- the
most conservative value across all workloads --- is \texttt{Xsafe}=6 for FP16,
\texttt{Xsafe}=4 for BF16, and \texttt{Xsafe}=15 for FP32, set in each case
by the most sensitive model class (text-conditional diffusion for FP16 and
BF16; PixArt-$\alpha$ for FP32). These floors are the hardware design
parameter: bits below \texttt{Xsafe} can be left unprotected; bits above it,
including the remaining upper fraction bits, require ECC coverage.

\paragraph{Stability of the floors.}
A natural question is whether the floors FP16=6, BF16=4, FP32=15 are
artifacts of the specific evaluation protocol.
Table~\ref{tab:xsafe-stability} addresses this by varying five axes of the
protocol and recording the effect on \texttt{Xsafe}. The floors are preserved
under the operating quality bound and are insensitive to sample count.
Varying dataset, injection site, and GPU architecture moves per-model
\texttt{Xsafe} by at most one bit, and the resilient models swept in these
tests stay well above the floor in every case. The fundamental reason the
floors are stable to quality bound choice is that the diffusion sensitivity
transition is near-vertical: SD3.5-Medium BF16 PSNR falls from 31\,dB to
14\,dB across just four adjacent bit positions, so the exact placement of the
quality threshold line does not change which bit position it selects. A
stricter bound tightens the floors monotonically as expected; a looser bound
relaxes them. Per-model \texttt{Xsafe} values should therefore be understood
as exhibiting bounded sub-bit drift across protocol variants, not literal
invariance. Stochastic behavior at realistic fault rates is analyzed in
\S~\ref{sec:ber-sweeps}.

\begin{table}[t]
\centering
\footnotesize
\setlength{\tabcolsep}{4pt}
\renewcommand{\arraystretch}{1.0}
\caption{\texttt{Xsafe}-floor stability across the evaluation protocol. Under
the operating quality bound the conservative floors (FP16=6, BF16=4,
FP32=15) are preserved; a stricter bound only tightens them. Per-model
\texttt{Xsafe} varies by at most one bit across datasets, sample counts, and
injection sites, and the swept resilient models stay well above the floor.}
\label{tab:xsafe-stability}
\begin{tabularx}{\linewidth}{@{}L{0.18\linewidth} L{0.42\linewidth} Y@{}}
\toprule
\textbf{Axis} & \textbf{Variation tested} & \textbf{Effect on \texttt{Xsafe}} \\
\midrule
Quality bound & Cliff 3/5\,dB; PSNR $-3$/$-6$\,dB & FP16=6, BF16=4 preserved \\
Quality bound & Stricter PSNR $-1$\,dB / SSIM bound & Tightens monotonically \\
Dataset/task & SST-2/MNLI/QQP; test-clean/other; ImageNet & $\leq$1 bit; above floor \\
Sample count & 500--5000 & 0 bit \\
Injection site & Attention vs.\ MLP/conv & $\leq$1 bit \\
\bottomrule
\end{tabularx}

\end{table}

\begin{table}[t]
\centering
\scriptsize
\setlength{\tabcolsep}{2.5pt}
\renewcommand{\arraystretch}{1.04}
\caption{Per-model \texttt{Xsafe} --- the number of fraction LSBs that can
safely remain unprotected without crossing the modality-specific quality bound
--- for FP16, BF16, and FP32 across 16 evaluated models. The cross-model floor (bold) is the conservative
hardware design parameter: FP16 \texttt{Xsafe}=6, BF16 \texttt{Xsafe}=4,
FP32 \texttt{Xsafe}=15. Floors are set by text-conditional diffusion models
for FP16 and BF16, and by PixArt-$\alpha$ for FP32; other families
tolerate wider unprotected regions. PPL = perplexity.}
\label{tab:cross-arch-xsafe}
\begin{tabularx}{\linewidth}{@{}L{0.22\linewidth} L{0.18\linewidth} C{0.08\linewidth} C{0.08\linewidth} C{0.08\linewidth} Y@{}}
\toprule
\textbf{Model} & \textbf{Family} & \textbf{FP16} & \textbf{BF16} & \textbf{FP32} & \textbf{Metric} \\
\midrule
\multicolumn{6}{@{}l}{\textit{Diffusion}} \\
DiT-XL/2 (675M) & Class-cond MHA & 7 & 6 & 20 & PSNR \\
SD 1.5 (860M) & Text-cond MHA+Cross & \textbf{6} & \textbf{4} & 18 & PSNR \\
PixArt-\(\alpha\) (611M) & Text-cond MHA+Cross & 8 & 5 & \textbf{15} & PSNR \\
SD3.5-Medium (2.5B) & MMDiT & \textbf{6} & \textbf{4} & OOM & PSNR \\
\midrule
\multicolumn{6}{@{}l}{\textit{Decoder LLMs}} \\
Llama-3 8B & GQA & 9 & 7 & 23 & PPL \\
Phi-3-mini (3.8B) & MHA & 10 & 7 & 23 & PPL \\
Mistral-7B & GQA+SWA & 10 & 7 & 24 & PPL \\
GPT-2 Medium (355M) & MHA & 9 & 6 & 22 & PPL \\
Falcon-7B & MQA (71:1) & 10 & 7 & 23 & PPL \\
\midrule
\multicolumn{6}{@{}l}{\textit{Encoder / Enc-Dec}} \\
BERT-base (110M) & MHA (bidir) & 12 & 8 & 25 & Accuracy \\
Whisper-small (244M) & MHA+Cross & 9 & 6 & 24 & WER \\
\midrule
\multicolumn{6}{@{}l}{\textit{Vision Encoders}} \\
ViT-L/16 (304M) & MHA & 11 & 8 & 24 & Top-1 \\
Swin-B (88M) & Shifted window & 14 & 8 & 27 & Top-1 \\
\midrule
\multicolumn{6}{@{}l}{\textit{CNNs (no self-attention)}} \\
ResNet-50 (25M) & None & 11 & 8 & 24 & Top-1 \\
EfficientNet-B0 (5.3M) & SE (channel) & 11 & 8 & 24 & Top-1 \\
MobileNetV3-L (5.4M) & SE (channel) & 10 & 7 & 23 & Top-1 \\
\midrule
\multicolumn{6}{@{}l}{\textit{Cross-model floor}} \\
\textbf{Minimum \texttt{Xsafe}} & & \textbf{6} & \textbf{4} & \textbf{15} & \\
\bottomrule
\end{tabularx}

\end{table}
\paragraph{Is \texttt{Xsafe} governed by model architecture?}
Figure~\ref{fig:llm_xsafe_vs_heads_kv} tests whether \texttt{Xsafe} variation
across decoder LLMs reflects attention head structure rather than data type, by
plotting \texttt{Xsafe} against the heads/KV ratio
($\texttt{num\_heads}/\texttt{num\_kv\_heads}$) for the five decoder LLMs ---
a ratio that increases under grouped-query and multi-query attention and could
in principle confer KV-cache redundancy that masks bit errors. Spearman
correlation is non-significant across all data types ($\rho{\approx}-0.031$,
$p{\approx}0.953$): attention head structure is not a predictor of
\texttt{Xsafe}. \texttt{Xsafe} is determined by the floating-point data type,
not by the model's attention configuration.
\begin{figure}[t]
  \centering
  \includegraphics[width=0.75\linewidth]{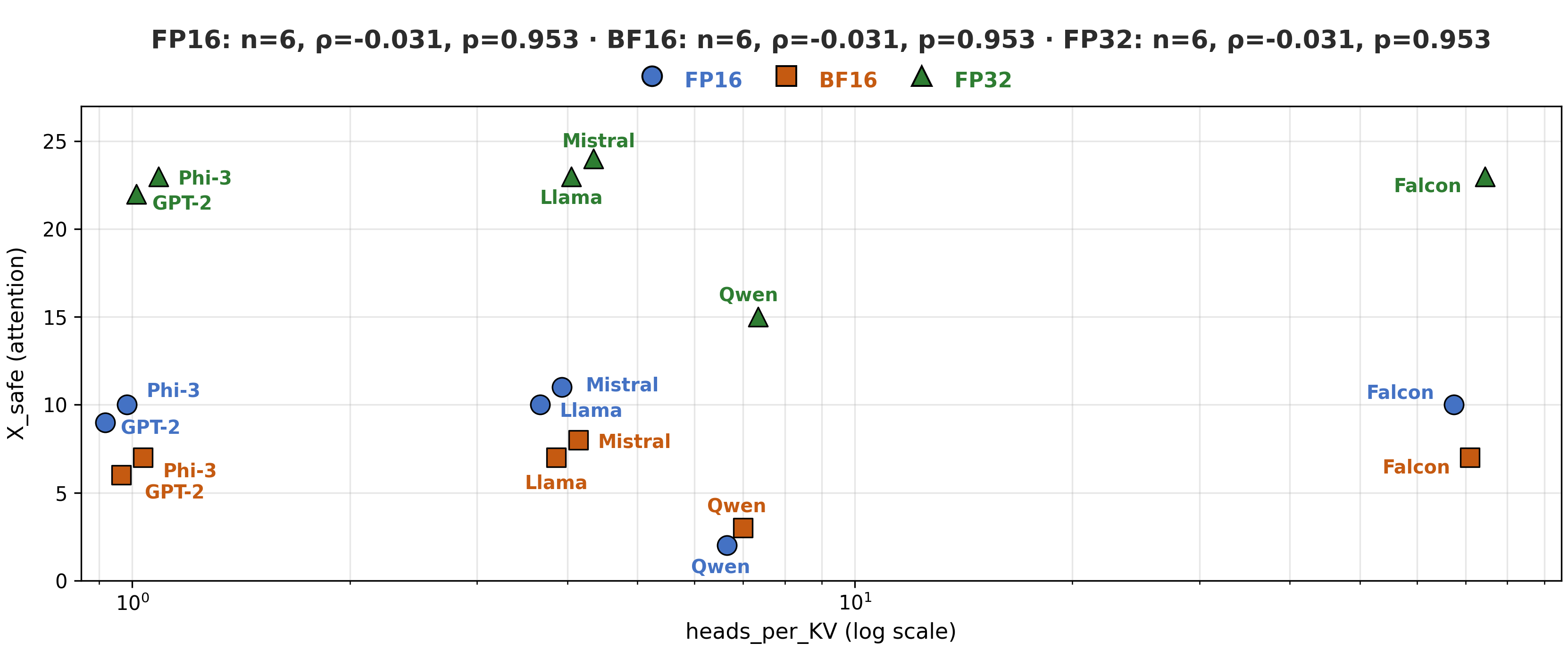}
  \caption{Attention-path \texttt{Xsafe} for five decoder LLMs plotted against
  their heads/KV ratio --- the number of query heads sharing each key-value
  head pair --- on a log scale, for FP16, BF16, and FP32. A higher heads/KV
  ratio (as in multi-query attention) compresses key-value memory across more
  query heads and could in principle confer redundancy that masks bit errors.
  Spearman correlation is non-significant across all data types
  ($\rho{\approx}-0.031$, $p{\approx}0.953$), ruling out attention head
  structure as a predictor of \texttt{Xsafe}.}
  \label{fig:llm_xsafe_vs_heads_kv}
\end{figure}
We therefore use \texttt{Xsafe}=6 (FP16), \texttt{Xsafe}=4 (BF16), and
\texttt{Xsafe}=15 (FP32), then derive workload-aware tiers in
\S~\ref{sec:tiers}.

\begin{table}[t]
\centering
\scriptsize
\setlength{\tabcolsep}{2pt}
\renewcommand{\arraystretch}{1.04}
\caption{Sharpness of the sensitivity transition near each model's
\texttt{Xsafe} boundary (FP16, single-bit-position faults at 1\% element
rate). Columns Bit~N through Bit~N+2 show a three-bit window starting at the
first bit above \texttt{Xsafe}: $N{=}10$ for Llama-3~8B and ViT-L/16,
$N{=}12$ for BERT-base, and $N{=}8$ for SD3.5-Medium. In every case, one or
two bits above the threshold move behavior from tolerable to catastrophic with
no gradual plateau, confirming \texttt{Xsafe} is a hard boundary.}
\label{tab:margin-to-failure}
\begin{tabularx}{\linewidth}{@{}L{0.16\linewidth} C{0.07\linewidth} C{0.09\linewidth} C{0.13\linewidth} C{0.13\linewidth} C{0.13\linewidth} Y@{}}
\toprule
Model & Xsafe & Baseline & Bit N & Bit N+1 & Bit N+2 & Metric (\(\Delta\)) \\
\midrule
Llama-3 8B & 9 & 6.64 PPL & 6.81 (b10) & 7.01 (b11) & 98.6 (b12) & PPL (\(\times\)14.8) \\
ViT-L/16 & 11 & 80.20\% & 80.34 (b10) & 76.26 (b11) & 73.38 (b12) & Top-1 (\(-\)3.94pp) \\
BERT-base & 12 & 92.43\% & 91.06 (b12) & 90.94 (b13) & 91.86 (b14) & Acc (\(-\)1.49pp) \\
SD3.5-Med & 6 & 1.0 & 0.950 (b6) & 0.873 (b7) & 0.568 (b8) & SSIM (\(-\)0.19/step) \\
\bottomrule
\end{tabularx}

\end{table}
The sensitivity pattern is also consistent across data types:
Table~\ref{tab:margin-to-failure} shows that running the same models in FP16,
BF16, and FP32 produces the same abrupt two-phase transition in each case, with
\texttt{Xsafe} shifting with the format at a model-dependent offset rather than
at a fixed bit position. Attention-free CNN classifiers reinforce this: they
match the data-type-scaled pattern of transformer models, confirming that the
sensitivity structure follows the floating-point format, not the attention
mechanism. Statistical confirmation follows.\label{sec:datatype-property}

\paragraph{Statistical confirmation of the boundary.}
Paired McNemar tests applied per bit position to BERT-base and ViT-L/16 ---
across attention and MLP paths for all three data types --- confirm the
transition statistically (Figure~\ref{fig:mcnemar_significant_ranges}). After
Bonferroni correction, significance is concentrated exclusively in the exponent
and sign positions; fraction LSBs are indistinguishable from baseline across all
six panels. The statistical scope is limited to these two classification models;
the deterministic sweeps in Table~\ref{tab:cross-arch-xsafe} and
Table~\ref{tab:margin-to-failure} extend the pattern to the remaining 14
models.
\begin{figure}[t]
  \centering
  \includegraphics[width=0.75\linewidth]{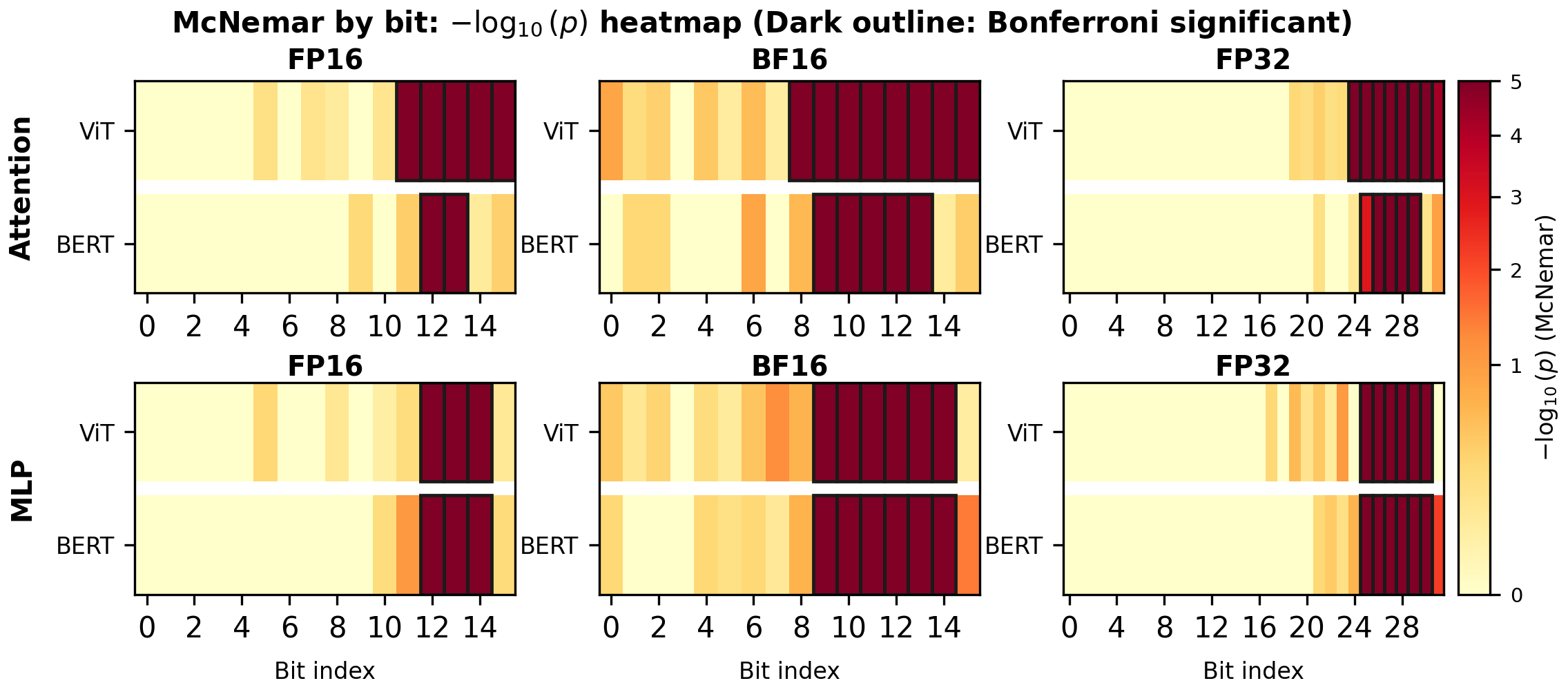}
  \caption{Paired McNemar tests for single-bit flips at matrix-multiply
  outputs (BERT-base SST-2, ViT-L/16 ImageNet), across attention and MLP
  paths, for FP16, BF16, and FP32. Color: $-\log_{10}(p)$ (darker = stronger
  significance); dark outlines mark Bonferroni-significant positions. Fraction
  LSBs are statistically indistinguishable from baseline across all panels;
  significance peaks at the exponent boundary (shifting with data type),
  consistently across attention and MLP paths.}
  \label{fig:mcnemar_significant_ranges}
\end{figure}

\subsubsection{Persistent-Fault Polarity Asymmetry}\label{sec:stuck-at}

\S~\ref{sec:xsafe-threshold} established \texttt{Xsafe} under transient
faults --- bits that flip and recover. We now relax this assumption and ask
whether \texttt{Xsafe} holds when a bit is permanently stuck at a fixed value,
as occurs under aging, retention failure, or manufacturing defects.
Figure~\ref{fig:stuck-at-polarity} sweeps stuck-at-0 and stuck-at-1 faults
across all bit positions for BERT-base, ViT-L/16, and Whisper-small, across
both attention and MLP paths and all three data types. \texttt{Xsafe}
survives: fraction LSBs remain benign under both polarities across all models
and data types, and the protection boundary is preserved.

However, persistent faults introduce a polarity asymmetry absent in transient
flips --- stuck-at-0 and stuck-at-1 cause qualitatively different damage at
the same bit position. On the sign bit, stuck-at-0 forces all values positive,
which can raise perplexity by more than 1500$\times$; stuck-at-1 at the same
position is far milder ($<\!1.1\times$) because attention scores are typically
near-zero or negative. Exponent bits show rate-dependent behavior: bit~10
stuck-at-1 in FP16 raises perplexity by only 1.05$\times$ when 0.1\% of words
are affected, but by 30$\times$ when 5\% of words are affected, because a
higher fraction of corrupted exponent bits pushes more values toward overflow.
Stuck-at-0 at the same position remains benign across all tested rates. The
same polarity pattern holds consistently across models and data types.

\begin{figure}[t]
\centering
\includegraphics[width=0.75\linewidth]{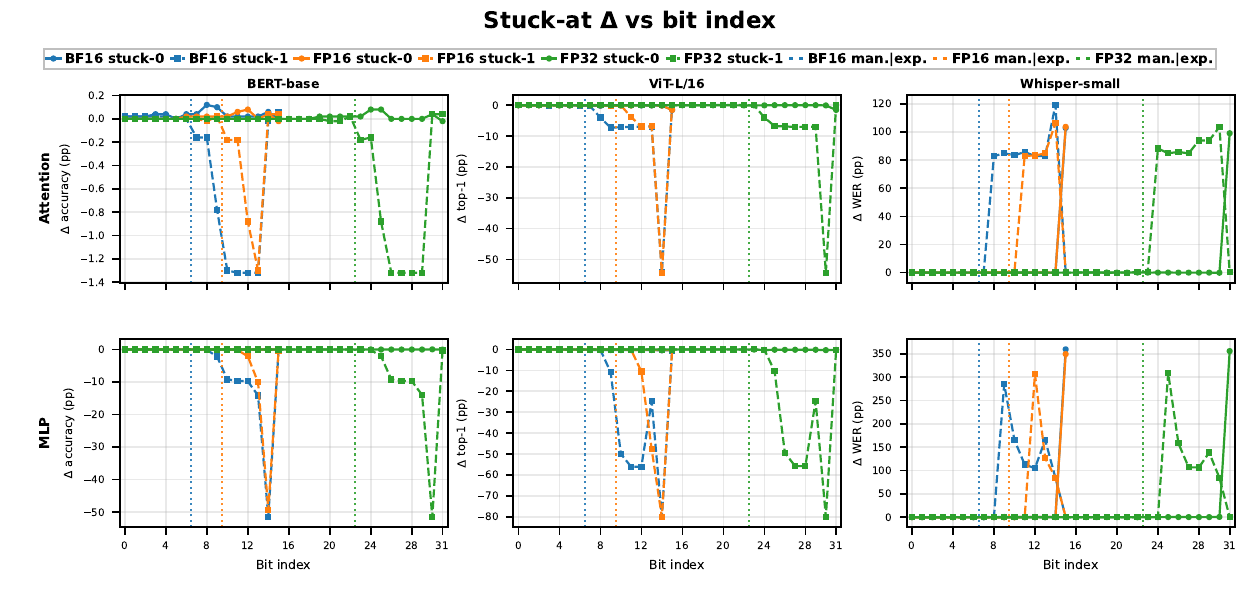}
\caption{Permanent stuck-at-0 (solid lines) and stuck-at-1 (dashed lines)
faults swept across all bit positions (1\% incidence) for BERT-base, ViT-L/16,
and Whisper-small, on attention (top) and MLP (bottom) paths. The $y$-axis is
the change in each model's primary quality metric; dotted lines mark the
fraction/exponent boundary per data type. Fraction LSBs remain benign under both
polarities, confirming \texttt{Xsafe} holds under persistent faults. Above the boundary, stuck-at-0
and stuck-at-1 produce asymmetric damage: stuck-at-1 at exponent bits inflates
magnitude toward overflow, while stuck-at-0 at the same position is typically
benign.}
\label{fig:stuck-at-polarity}
\end{figure}

\subsubsection{Stochastic BER Sweeps}\label{sec:ber-sweeps}

\S~\ref{sec:xsafe-threshold} used deterministic sweeps that stress one fixed
bit position at a time across every word in the targeted tensor --- a
controlled experiment that isolates bit importance but does not reflect how
faults occur in real hardware, where any bit in a word may flip at random. We
now relax both the deterministic and the uniform-fault assumptions and ask
whether \texttt{Xsafe} holds under realistic stochastic fault conditions.

\paragraph{Rate dependence.}
Figure~\ref{fig:threshold-site-summary} plots aggregate PSNR for SD3.5-Medium
as a function of the per-word fault rate $p_\text{word}$ --- the probability
that any given floating-point word suffers a random bit-flip --- across the
range $10^{-6}$ to $10^{-1}$. Quality remains above 24\,dB for $p_\text{word}
\leq 10^{-4}$ and collapses catastrophically near $10^{-3}$, producing a
sharp usable-to-catastrophic threshold rather than a gradual degradation curve.
The inset breaks this down by bit class at three fault rates: even at the high
fault rate of $p_\text{word} = 10^{-2}$, LSB mantissa bits remain near
27\,dB while sign and exponent bit corruption collapses PSNR to below 10\,dB.
This confirms that bit class --- and therefore \texttt{Xsafe} --- remains the
dominant determinant of fault severity at realistic rates, not fault rate
alone. Across all 16 evaluated models, degradation stays within 1\% for
$p_\text{word} \leq 10^{-4}$ and becomes rapid near $10^{-3}$.

\begin{figure}[t]
\centering
\includegraphics[width=0.85\linewidth]{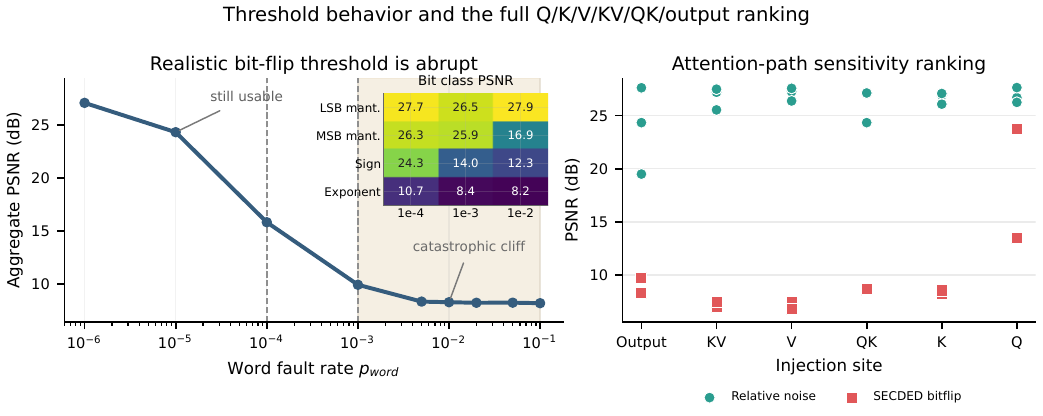}
\caption{Aggregate SD3.5-Medium PSNR under random bit-flips as a function of
per-word fault rate $p_\text{word}$. Quality stays above 24\,dB for
$p_\text{word} \leq 10^{-4}$ and collapses near $10^{-3}$. The inset breaks PSNR
down by bit class at $10^{-4}$/$10^{-3}$/$10^{-2}$: LSB mantissa bits stay near
27\,dB even at high rates, while sign/exponent corruption drops below 10\,dB.
Bit class --- not fault rate alone --- is the dominant determinant of fault
severity.}
\label{fig:threshold-site-summary}
\end{figure}

\subsubsection{Per-Layer and Per-Modality Sensitivity Tiers}\label{sec:tiers}\label{sec:layer_sensitivity}

The conservative \texttt{Xsafe} floor guarantees safety across all workloads
but is set by the most sensitive model. We now ask whether it can be safely
widened for more resilient workloads without retraining.

\paragraph{Per-layer sensitivity spread.}
Gaussian noise injected one layer at a time reveals that sensitivity varies
substantially within a single model (Table~\ref{tab:per-layer-metric-spread-tiers}):
the spread between the most and least sensitive layers exceeds an order of
magnitude for LLMs and spans 3--14\,dB for diffusion models. Uniform
per-layer protection therefore simultaneously over-provisions resilient layers
and under-provisions critical ones. This wide intra-model spread motivates a
tiered protection policy where the bypass width can be adjusted per workload.

\begin{table}[htbp]
\centering
\footnotesize
\setlength{\tabcolsep}{3pt}
\renewcommand{\arraystretch}{1.0}
\caption{Per-layer sensitivity spread under Gaussian noise injection
($\sigma{=}0.01$, one layer at a time) for 8 core evaluation models plus 3
auxiliary diffusion checkpoints. The remaining 8 evaluation models were not
profiled at this granularity; their per-model \texttt{Xsafe} values appear in
Table~\ref{tab:cross-arch-xsafe}. Spreads exceed an order of magnitude for
LLMs and 3--14\,dB for diffusion.}
\label{tab:per-layer-metric-spread-tiers}
\begin{tabularx}{\linewidth}{@{}L{0.22\linewidth} L{0.20\linewidth} C{0.12\linewidth} C{0.14\linewidth} C{0.16\linewidth}@{}}
\toprule
Model & Metric & Min & Max & Spread \\
\midrule
DiT-XL/2 & PSNR (dB) & 16.55 & 30.24 & 13.69 \\
SD 1.5 & PSNR (dB) & 20.85 & 26.34 & 5.49 \\
PixArt-$\alpha$ & PSNR (dB) & 18.15 & 30.15 & 12.00 \\
SDXL & PSNR (dB) & 30.10 & 34.17 & 4.07 \\
Llama-3 8B & PPL ratio & 12.5$\times$ & 174,520$\times$ & 174,507$\times$ \\
Falcon-7B & PPL ratio & 1.05$\times$ & 189.9$\times$ & 188.9$\times$ \\
BERT-base & Acc.\ delta (pp) & $-$8.03 & $-$0.34 & 7.68 \\
Whisper-small & WER delta (pp) & $-$0.64 & $+$86.90 & 87.54 \\
ViT-L/16 & Logit cos.\ div. & 0.0000 & 0.0467 & 0.0467 \\
Playground V2.5 & PSNR (dB) & 23.13 & 27.40 & 4.27 \\
SDXL Turbo & PSNR (dB) & 15.11 & 18.39 & 3.28 \\
\bottomrule
\end{tabularx}

\end{table}

\paragraph{Per-modality \texttt{Xsafe} tiers.}
Grouping the per-model \texttt{Xsafe} values from
Table~\ref{tab:cross-arch-xsafe} by modality yields three natural tiers
(Table~\ref{tab:fp16-modality-protection-tiers}), covering all workloads from
the most conservative (text-conditional diffusion, Tier~0, \texttt{Xsafe}=6)
through resilient vision, NLU, and CNN workloads (Tier~2, \texttt{Xsafe}=10).
A runtime-configurable bypass width allows a single
hardware design to serve all three tiers without redesign or retraining; the
hardware implications are developed in \S~\ref{sec:architecture}.

\begin{table}[htbp]
\centering
\footnotesize
\setlength{\tabcolsep}{3pt}
\renewcommand{\arraystretch}{1.0}
\caption{Three FP16 protection tiers with their \texttt{X} value, modality
coverage, unprotected bit count, and cumulative model coverage. Tier
boundaries use family minima, ensuring no model is exposed beyond its
validated \texttt{Xsafe} threshold.}
\label{tab:fp16-modality-protection-tiers}
\begin{tabularx}{\linewidth}{@{}C{0.04\linewidth} C{0.06\linewidth} L{0.18\linewidth} L{0.36\linewidth} C{0.10\linewidth} C{0.10\linewidth}@{}}
\toprule
Tier & \texttt{X} & Modality & Models covered & Unprotected bits (FP16) & $\ \ $ Models in tier \\
\midrule
0 & 6 & All (conservative floor incl.\ text-cond.\ diffusion) &
SD3.5-Medium, PixArt-\(\alpha\), SD 1.5, DiT-XL/2, + all below & 6 of 16
& 4 \\
1 & 9 & LLMs, enc-dec & Llama-3 8B, GPT-2 Medium, Whisper-small, +
all below & 9 of 16 & 7 \\
2 & 10 & Vision, NLU, resilient LLMs, CNNs & Phi-3-mini,
Mistral-7B, Falcon-7B, MobileNetV3-L, ViT-L/16, ResNet-50, EfficientNet-B0, BERT-base, Swin-B & 10 of 16 & 16 \\
\bottomrule
\end{tabularx}

\end{table}

\subsubsection{Multi-Cell Upset Robustness}\label{sec:mcu}

All preceding subsections injected single-bit faults. We now relax this
assumption: in advanced process nodes, a single particle strike can flip
multiple physically adjacent memory cells simultaneously, producing a
multi-cell upset of $N$ contiguous bits (MCU-$N$)~\cite{ibe2010mcu}. We ask
whether \texttt{Xsafe} holds when an event corrupts 2 or 3 adjacent bits
rather than one.
MCU-$N$ experiments repeat the deterministic sweeps of
\S~\ref{sec:xsafe-threshold} but corrupt $N$ contiguous bits per event.
Table~\ref{tab:mcu-xsafe-contraction} reports the results: \texttt{Xsafe}
contracts by at most one bit per additional flipped bit across all tested data
types and architectures. Figure~\ref{fig:mcu-summary} confirms this
contraction is consistent across models. The contraction is bounded because an
MCU-$N$ event whose entire footprint lies within the safe LSB region causes no
additional damage; only events that straddle the \texttt{Xsafe} boundary
produce a contraction, and in that case the outcome is dominated by the
highest corrupted bit position --- equivalent in effect to a single-bit fault
at that position. The $N{-}1$ contraction is therefore a worst case arising
only at the boundary, not a general property of all MCU events.

\begin{figure}[t]
\centering
\includegraphics[width=0.75\linewidth]{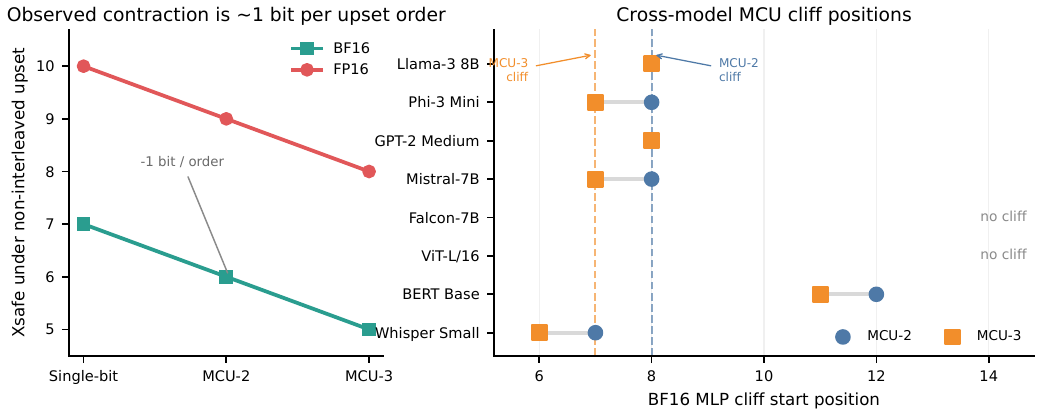}
\caption{Left: BF16 and FP16 \texttt{Xsafe} under MCU-2 and MCU-3 compared
to single-bit faults, showing a contraction of at most one bit per additional
flipped bit. Right: cross-architecture BF16 MLP sensitivity-transition
positions under MCU-2 and MCU-3, confirming that the contraction pattern is
consistent across all evaluated models.}
\label{fig:mcu-summary}
\end{figure}

\begin{table}[t]
\centering
\footnotesize
\caption{Cross-data-type \texttt{Xsafe} contraction under MCU-2 and MCU-3.
Each entry reports the worst-case safe-LSB count across all evaluated models.
\texttt{Xsafe} contracts by at most one bit per additional flipped bit ---
MCU-2 costs at most one safe bit, MCU-3 costs at most two --- and only when
the upset footprint straddles the \texttt{Xsafe} boundary.}
\label{tab:mcu-xsafe-contraction}
\begin{tabular}{lccc}
  \toprule
  & Single-bit & MCU-2 & MCU-3 \\
  \midrule
  FP16 & $\texttt{Xsafe}=6$ & $\texttt{Xsafe}\geq5$ & $\texttt{Xsafe}\geq4$ \\
  BF16 & $\texttt{Xsafe}=4$ & $\texttt{Xsafe}\geq3$ & $\texttt{Xsafe}\geq2$ \\
  FP32 & $\texttt{Xsafe}=15$ & $\texttt{Xsafe}\geq14$ & $\texttt{Xsafe}\geq13$ \\
  \bottomrule
\end{tabular}

\end{table}

\subsection{Continuous Noise Tolerance}\label{sec:continuous-results}

\S~\ref{sec:discrete-results} characterized sensitivity to discrete digital
bit-flips. We now ask whether the same conclusions hold under continuous analog
noise --- perturbations that nudge floating-point values by some amount rather
than flipping a specific bit, as occurs under voltage scaling, thermal
fluctuation, or device variation. The central question is: does noise
distribution shape alone determine fault severity, or does the way noise
couples to the signal matter more?

\paragraph{Coupling mode dominates distribution shape.}
Four noise families --- Gaussian, log-normal, log-uniform, and
matched-variance bit-flip --- are each injected under the two coupling modes
defined in \S~\ref{fault-model}, across 3{,}780 images on SD3.5-Medium in
FP16, BF16, and FP32. The dominant effect is coupling mode, not distribution
shape: relative coupling yields approximately 22\,dB higher PSNR than constant
coupling across all noise families. Within each coupling mode, severity orders
as Gaussian $>$ log-uniform $>$ log-normal $\gg$ bit-flip
(Figure~\ref{fig:shape-matters-bridge}). The critical observation is that
digital bit-flips are the most damaging family even at matched variance --- a
28--40\,dB gap relative to Gaussian noise at identical variance --- confirming
that fault severity is determined by which bit fields are corrupted, not by
noise magnitude. The same coupling-mode ordering holds on MLP blocks, with
only a 1--3\,dB absolute PSNR shift.

\begin{figure}[t]
\centering
\includegraphics[width=0.75\linewidth]{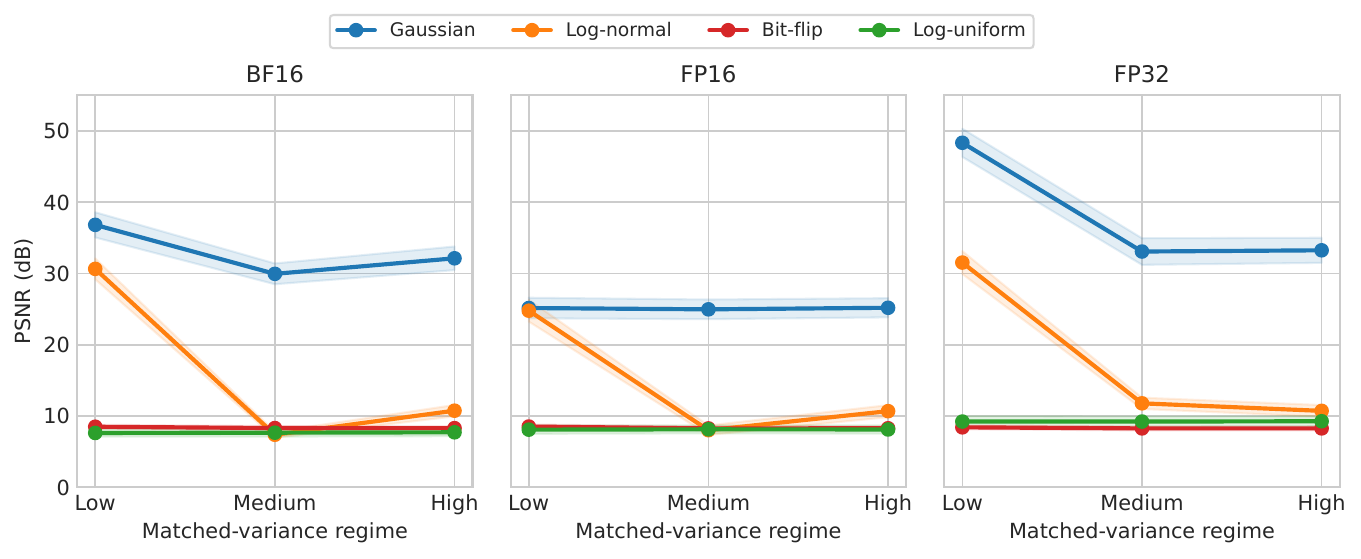}
\caption{Matched-variance comparison across four noise families (Gaussian,
log-normal, bit-flip, log-uniform) at low, medium, and high variance regimes,
for BF16, FP16, and FP32. At identical variance, digital bit-flips (red) are
28--40\,dB worse than Gaussian noise (blue), confirming that fault severity is
determined by which bit fields are corrupted, not by noise magnitude alone.}
\label{fig:shape-matters-bridge}
\end{figure}

\paragraph{Generalization across diffusion checkpoints.}
The continuous-noise campaign above is run on SD3.5-Medium only.
Figure~\ref{fig:diffusion-harness-summary-heatmap} confirms the severity
ordering generalizes across additional diffusion checkpoints, with no
checkpoint showing a qualitatively different pattern.

\begin{figure}[t]
\centering
\includegraphics[width=0.75\linewidth]{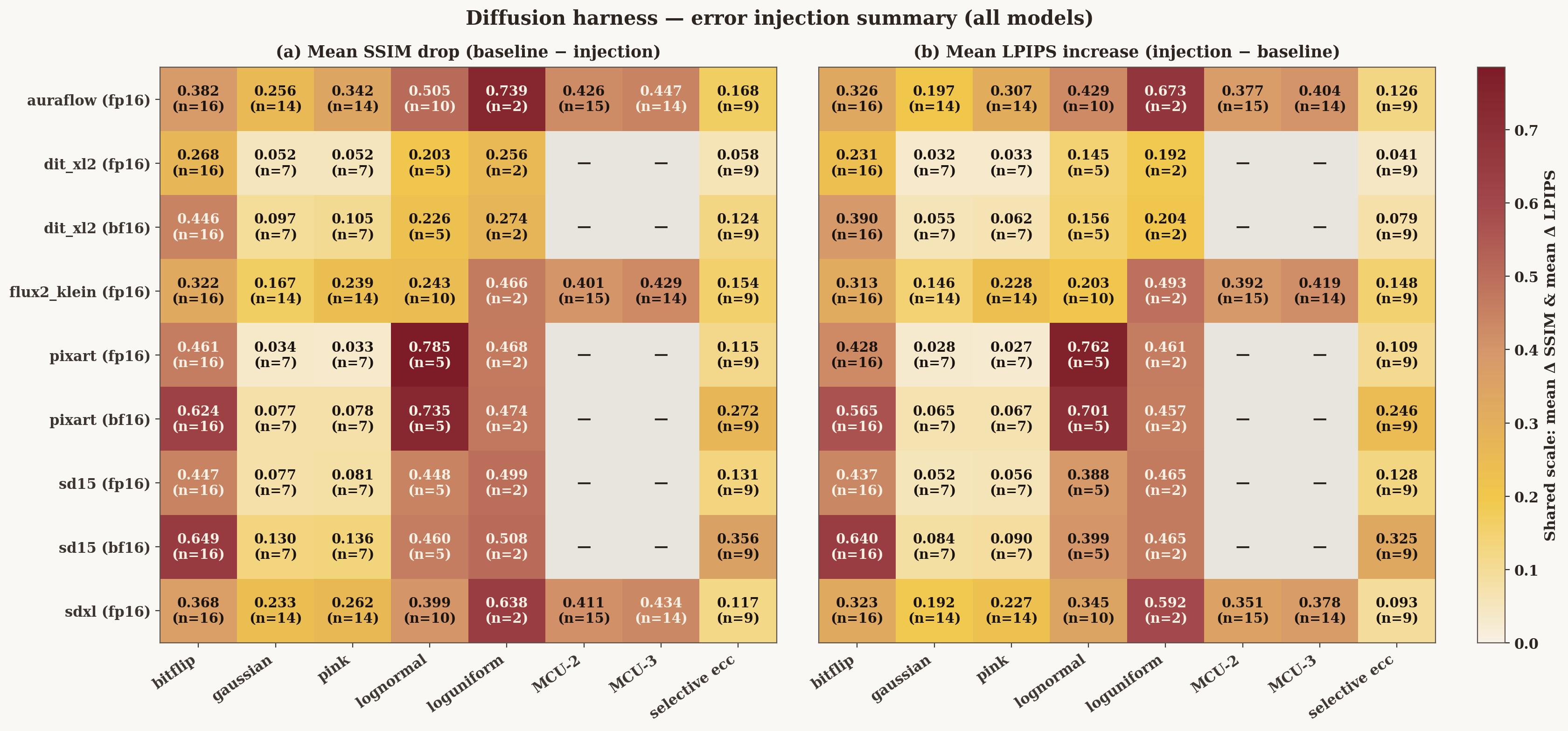}
\caption{Cross-model diffusion harness summary showing mean SSIM drop (a) and
mean LPIPS increase (b) relative to clean baselines, across evaluated
checkpoints (rows) and experimental conditions (columns). The severity
ordering of noise families is consistent across all checkpoints, confirming
that the coupling-mode and distribution-shape findings from SD3.5-Medium
generalize across diffusion model architectures.}
\label{fig:diffusion-harness-summary-heatmap}
\end{figure}

} 

\section{Hardware Architecture: Unequal Error Protection for ML Inference}\label{sec:architecture}
\rev{\S~\ref{sec:discrete-results} and \S~\ref{sec:continuous-results} converge on
three findings that constrain the hardware design. First, \texttt{Xsafe} --- a data-type-specific
threshold within the fraction field --- defines which bits can safely remain
unprotected: FP16 \texttt{Xsafe}=6, BF16 \texttt{Xsafe}=4, FP32
\texttt{Xsafe}=15, with workload-aware tiers that widen the unprotected region
for more resilient modalities (\S~\ref{sec:tiers}). Second, digital bit-flips
degrade quality by 28--40\,dB more than Gaussian noise of identical variance
(Figure~\ref{fig:shape-matters-bridge}), confirming that any viable protection
scheme must distinguish \emph{which} bits are corrupted, not merely \emph{how
many} --- this is unequal error protection (UEP) by definition. Third,
multi-cell upsets contract \texttt{Xsafe} by at most one bit per additional
flipped bit, and only when the upset footprint straddles the protection
boundary (\S~\ref{sec:mcu}). These three findings are the inputs to the
hardware design that follows.}

Translating this into hardware imposes four requirements:
\rev{
\begin{enumerate}[topsep=0pt, leftmargin=*, label=\textbf{R\arabic*}]
\item \label{req:selective} \emph{Selective application:} UEP applies only to
inference-produced floating-point values (activations, partials,
accumulators); weights, inputs, and integer control data retain full ECC.
Fused-attention kernels already separate weight traffic from result
traffic~\cite{dao2025flashattention4, modal2025reverseengineer}, so UEP
targets the result path without affecting inputs.
\item \label{req:metadata} \emph{Per-cacheline metadata:} each cacheline must
encode its active bypass width, because \texttt{Xsafe} varies by data type and
modality tier. A single 3-bit tag per cacheline suffices since ML output
tensors are uniformly typed.
\item \label{req:partition} \emph{Partitioned SRAM:} bypassed LSBs must be
physically separated from protected bits and check bits, so the non-critical
partition can operate at reduced voltage for energy savings proportional to
bypass width.
\item \label{req:config} \emph{Modality-aware configurability:} the bypass
width $X$ must be programmable at model load time to match the workload's
modality tier without hardware redesign.
\end{enumerate}}
\begin{figure}[t]
\centering
\includegraphics[width=0.55\linewidth]{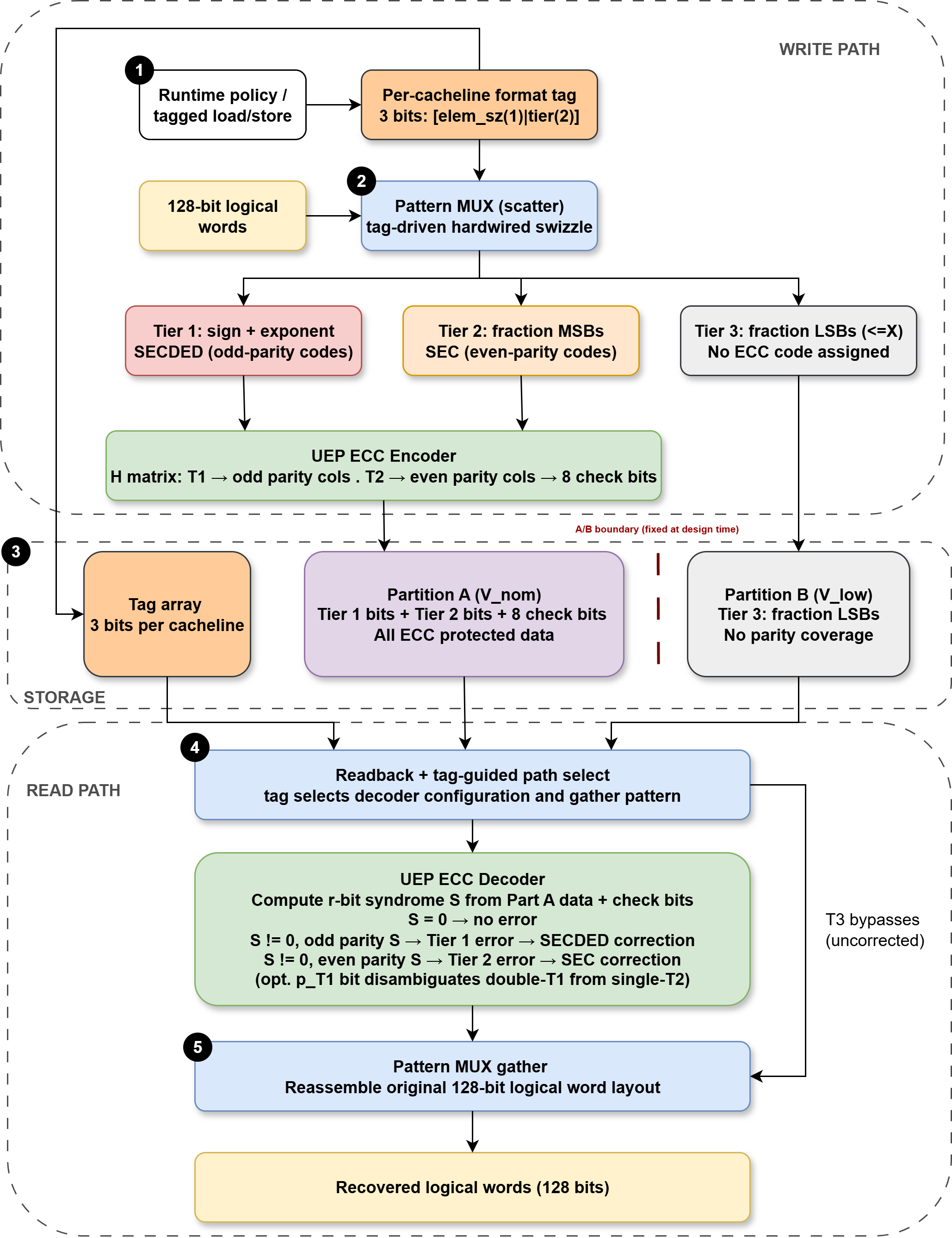}
\caption{Tagged selective-ECC access path for one 128-bit granule: the
per-cacheline tag selects routing and recovery (Pattern MUX/gather and
SECDED/SEC/bypass decode path), while the physical Partition~A/B boundary
remains fixed.}
\label{fig:selective-ecc-architecture}
\end{figure}
\noindent
The architecture satisfying R1--R4 has three components
(Figure~\ref{fig:selective-ecc-architecture}): a UEP codec
(\S~\ref{sec:uep_ecc_codec}) that computes check bits over the $N-X$
most-significant bits; \rev{a per-cacheline format tag (\S~\ref{sec:tag}) that
stores the active mode and is programmed at model load;} and a dual-partition
SRAM macro (\S~\ref{sec:sram}) that separates protected and bypassed bits,
\rev{with the bypass partition running at $V_{\mathrm{low}}$ for area and energy
savings.}

\rev{The per-data-type \texttt{Xsafe} floors (FP16=6, BF16=4, FP32=15) vary by at
most one bit across workloads, datasets, and injection sites
(Table~\ref{tab:xsafe-stability}). This stability is what makes it practical
to fix the protection boundary in hardware at design time rather than provision
it entirely at runtime. A purely runtime-configurable boundary would need to
accommodate the worst-case workload at every operating point, forfeiting the
energy savings that a fixed low-voltage partition enables. The per-cacheline
tag (R2, R4) preserves per-workload flexibility above this fixed floor:
software can widen the unprotected region for resilient workloads at model load
time without any hardware change.}

\subsection{UEP ECC Codec}\label{sec:uep_ecc_codec}
\subsubsection{Background: Hamming SECDED}
Hsiao SECDED~\cite{hsiao1970} protects \(N\) data bits with \(r\) parity bits
(\(r \approx \lceil\log_2(N)\rceil + 2\)). The code is defined by a
parity-check matrix \(\mathbf{H}\in\{0,1\}^{r\times(N+r)}\) whose columns are
unique bit-position codes. On a read, parity checks produce an \(r\)-bit
syndrome: zero means no error; non-zero identifies a single-bit error for
correction. In Hsiao SECDED, data-bit columns use odd-parity codes, so
single-bit errors produce odd syndromes and double-bit errors produce even
syndromes (detected, not corrected).
Among the \(2^r-1\) non-zero length-\(r\) codes, half are odd parity and half
are even parity. Standard SECDED uses only the odd-parity class for data bits,
leaving the even-parity class unused.

\subsubsection{Three-Tier Unequal Protection}
\label{subsec:three-tier-unequal-protection}
The UEP codec reduces ECC cost by observing that not all bit 
positions require the same level of protection. Our proposed 
three-tier UEP partitions the bits of each floating-point number 
into three tiers:

\begin{table}[htbp]
\centering
\footnotesize
\setlength{\tabcolsep}{4pt}
\renewcommand{\arraystretch}{1.0}
\caption{Three-tier partition of sign+exponent bits, protected 
fraction MSBs, and unprotected fraction LSBs for a packed 
codeword of \(W\) floating-point words.}
\label{tab:uep-tier-bit-partition}
\begin{tabularx}{\linewidth}{@{}C{0.11\linewidth} 
  C{0.3\linewidth} C{0.22\linewidth} C{0.22\linewidth} Y@{}}
\toprule
Tier & Bits Covered & Protection Level & Code Assignment \\
\midrule
1 & Sign + exponent & SECDED & Odd-parity codes \\
2 & Upper fraction (MSBs above bypass width $X$) & SEC 
  & Even-parity codes \\
3 & Lower fraction (LSBs within bypass width $X$) 
  & None --- bypassed & No code assigned \\
\bottomrule
\end{tabularx}

\end{table}

This separation provides two benefits:

\begin{enumerate}[topsep=0pt, leftmargin=*]
  \item \emph{Graded correction.} The decoder infers the tier
    directly from syndrome parity in one cycle and applies the
    corresponding recovery action (SECDED for Tier~1, SEC for
    Tier~2), with no additional lookup logic.

  \item \emph{Reduced check-bit demand.} Available codes per parity class are
    \(C_{\text{odd}} = 2^{r-1} - r\) (single-1 codes reserved for check-bit
    columns) and \(C_{\text{even}} = 2^{r-1} - 1\) (excluding all-zero). In
    standard SECDED, all \(N\) data bits compete for \(C_{\text{odd}}\). In
    UEP, Tier~1 and Tier~2 split across odd/even classes, which can reduce
    required \(r\).
\end{enumerate}

\subsubsection{Trade-Off: Reduced Double-Error Detection}
\label{sec:ded_tradeoff}

\begin{figure}[t]
  \centering
  \includegraphics[width=0.75\linewidth]{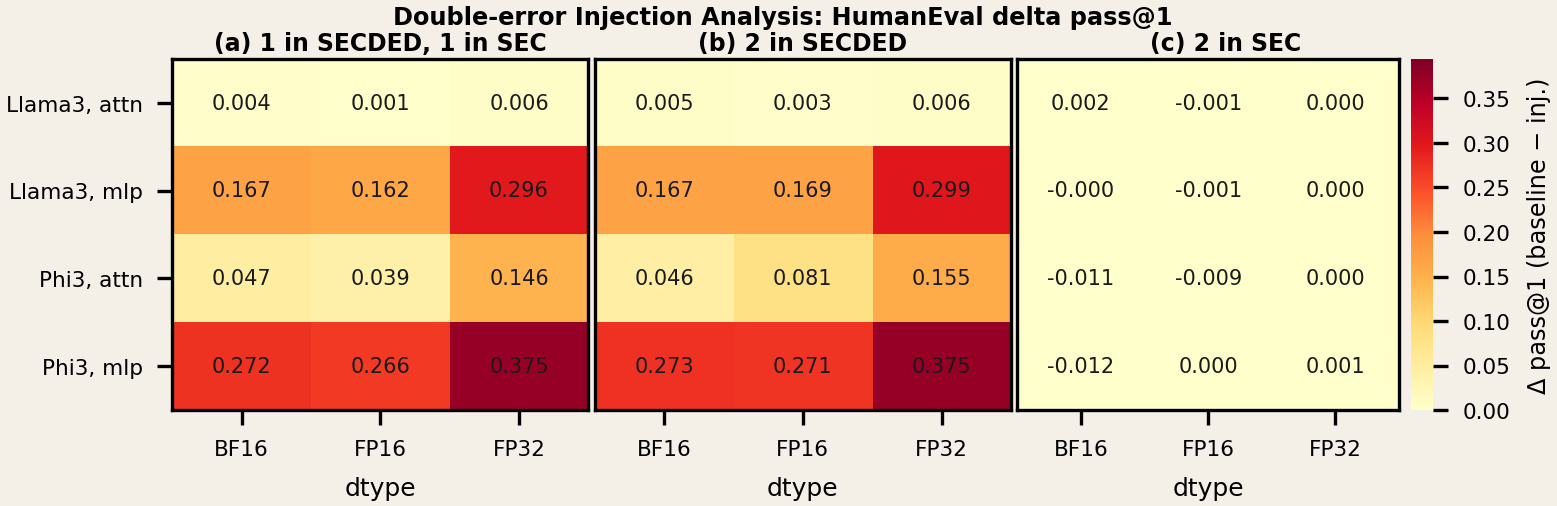}
  \caption{Mean HumanEval pass@1 drop (baseline minus injected), averaged over 32-layer sweeps, for three UEP double-error patterns and BF16/FP16/FP32, on Llama-3~8B and Phi-3-mini (attention vs.\ MLP).}
  \label{fig:uep_double_error_humaneval_dtype_heatmap}
\end{figure}

Assigning even-parity codes to Tier~2 breaks the SECDED parity invariant:
an even syndrome now admits two interpretations---\emph{(i)} a single Tier~2
error (correctable) or \emph{(ii)} a double Tier~1 error (uncorrectable). The
syndrome alone cannot distinguish them.
To stress this ambiguity, we inject three double-error patterns on two decoder
LLMs (Llama-3~8B and Phi-3-mini, full 32-layer sweeps). MMLU mean
\(\Delta\) remains \({\sim}10^{-3}\), while HumanEval shows larger pass@1
drops; aggregate attention-path impact is not catastrophic
(Figure~\ref{fig:uep_double_error_humaneval_dtype_heatmap}) even with three rare errors.

For \emph{transient} faults, ambiguity is usually benign: two independent
same-cycle errors in one codeword scale as \(\mathrm{BER}^{2}\); at raw
\(\mathrm{BER}=10^{-7}\), the expected interval exceeds \(10^{14}\) hours.
Even rare miscorrections land in Tier~2 fraction bits, which are less
critical by design (\S\ref{sec:continuous-results}).
For \emph{persistent} faults, risk is higher: a stuck Tier~1 bit can combine
with a transient Tier~1 error to produce an even syndrome that miscorrects a
Tier~2 bit while leaving the stuck sign/exponent bit uncorrected.

This path is closed by adding an extra check-bit and restoring the DED guarantee by occupying only the odd columns. Safety-critical paths may enable this while throughput-oriented paths may omit it and accept the
\(\mathrm{BER}^{2}\) residual risk. Section
~\ref{design-space-exploration-and-reliability} quantifies FIT impact.

\subsubsection{Codeword Construction}
Multiple floating-point words are packed into a single codeword 
aligned to the SRAM word width. For example, eight BF16 words 
form a 128-bit data payload. The codec assigns Tier~1 and Tier~2
protected bits to columns in~$\mathbf{H}$. The lower~$X$ 
fraction bits per word are assigned no code and excluded from 
the codeword entirely --- they are stored in a separate SRAM 
with no parity coverage.

\subsubsection{Check-Bit Budget}\label{sec:check_bit_budget}
Given the capacities above, the minimum check-bit count \(r_{\min}\) is the
smallest \(r\) for which both parity classes accommodate their tier demands:
\begin{align}
  (S + E) \times W &\leq 2^{r-1} - r 
    &&\text{(Tier~1: odd-parity codes)} 
    \label{eq:tier1-constraint} \\
  (F - X) \times W &\leq 2^{r-1} - 1 
    &&\text{(Tier~2: even-parity codes)}
    \label{eq:tier2-constraint}
\end{align}
where \(S\), \(E\), and \(F\) are sign/exponent/fraction widths, \(X\) is
bypass width, and \(W\) is words per codeword. Table~\ref{tab:check-bit-budget}
evaluates \(r=7\) and \(r=8\). At \(r=7\), BF16 Tier~1 demand
\((1+8)\times 8=72\) exceeds the \(57\) available odd-parity codes, so
\(r=7\) is only viable for FP16/FP32. At \(r=8\), all three data types fit;
therefore BF16-capable systems use \(r=8\), giving a fixed
\(128+8=136\)-bit codeword across data types.
\begin{table}[t]
\centering
\footnotesize
\setlength{\tabcolsep}{4pt}
\renewcommand{\arraystretch}{0.95}
\caption{Check-bit budget at $r{=}7$ (left) and $r{=}8$ 
(right). $W$~is the number of floating-point elements per 
codeword. T1 and T2 refer to Tier~1 and Tier~2 demands. 
The \emph{Codes} row shows the number of 
available codes in each parity class ($C_{\text{odd}}$ and 
$C_{\text{even}}$). A demand exceeding the available codes 
is marked~\xmark. See 
Table~\ref{tab:uep-tier-bit-partition} for tier definitions.}
\label{tab:check-bit-budget}
\begin{minipage}{0.48\columnwidth}
\centering
\subcaption*{$r = 7$}
\begin{tabular}{l c c c}
\toprule
       & $W$ & T1 & T2 \\
\midrule
Codes  &     & 57 & 63 \\
\midrule
BF16   & 8   & 72\,\xmark & 16\,\cmark \\
FP16   & 8   & 48\,\cmark & 32\,\cmark \\
FP32   & 4   & 36\,\cmark & 32\,\cmark \\
\bottomrule
\end{tabular}
\end{minipage}
\hfill
\begin{minipage}{0.48\columnwidth}
\centering
\subcaption*{$r = 8$}
\begin{tabular}{l c c c}
\toprule
       & $W$ & T1  & T2  \\
\midrule
Codes  &     & 120 & 127 \\
\midrule
BF16   & 8   & 72\,\cmark & 16\,\cmark  \\
FP16   & 8   & 48\,\cmark & 32\,\cmark  \\
FP32   & 4   & 36\,\cmark & 32\,\cmark  \\
\bottomrule
\end{tabular}
\end{minipage}

\end{table}

\subsubsection{Implementation and area}
The UEP codec is implemented in synthesizable Verilog RTL and verified with
146{,}000+ directed and random vectors across all tiers, including double-bit
patterns for DED. Table~\ref{tab:codec_impl} summarizes synthesis results.
The 27.8\% area reduction follows from the reduced code-space requirement
(\S\ref{sec:check_bit_budget}), and cross-tier miscorrection MTBF supports the
code-separation argument in \S\ref{subsec:three-tier-unequal-protection}.
System-level energy/reliability composition is evaluated in
\S\ref{design-space-exploration-and-reliability}.

\begin{table}[t]
\footnotesize
\centering
\caption{UEP codec implementation summary (sky130, TT corner,
         1.8\,V, 25\,\textdegree C; 7\,nm scaled estimate in
         parentheses).}
\label{tab:codec_impl}
\begin{tabular}{@{}lr@{}}
\toprule
\textbf{Metric} & \textbf{Value} \\
\midrule
Encoder area          & 6{,}884\,$\mu$m$^2$ \\
Decoder area          & 11{,}266\,$\mu$m$^2$ \\
\textbf{Total codec area}
                      & \textbf{17{,}895\,$\mu$m$^2$} \\
Uniform-SECDED baseline
                      & 24{,}780\,$\mu$m$^2$ \\
\textbf{Area reduction vs.\ baseline}
                      & \textbf{27.8\%} \\
\midrule
BF16 sub-codec ($r{=}8$)
                      & 825\,$\mu$m$^2$ \\
\quad Detection coverage
                      & 87.5\% \\
\quad Correction coverage
                      & 68.75\% \\
\midrule
Full datapath (tag\,+\,MUX\,+\,codec)
                      & 2{,}220\,$\mu$m$^2$
                        (${\sim}$15\,$\mu$m$^2$@7\,nm) \\
\quad vs.\ 256${\times}$256 systolic tile
                      & ${<}\,0.004$\% \\
\midrule
Cross-tier miscorr.\ MTBF\rlap{$^{\dagger}$}
                      & \husdraft{${\sim}\,$900\,yr} \\
\bottomrule
\end{tabular}\\[2pt]
\raggedright\scriptsize
\husdraft{$^{\dagger}$\,$\lambda_{\mathrm{SEU}}{=}10^{-12}$/bit/hr,
BER${=}10^{-4}$, 8\,MB SRAM.}

\end{table}

\subsection{Per-Cacheline Format Tag}
\label{sec:tag}
The UEP codec (\S~\ref{sec:uep_ecc_codec}) assumes known data type and bypass width
\(X\) per 128-bit codeword. To satisfy Requirements~\ref{req:metadata} and
\ref{req:config}, we attach a lightweight per-cacheline format tag, inspired by
hardware memory-tagging schemes (ARM MTE~\cite{arm_mte_whitepaper},
lowRISC~\cite{lowrisc2014tagged}).

\subsubsection{Tag Encoding}
\label{sec:tag_encoding}
Each 128-byte cacheline carries a 3-bit tag stored alongside the
existing cache-state metadata. The encoding is:
\begin{itemize}[leftmargin=*, topsep=0pt]
  \item \texttt{tag[2]} — element size: \texttt{0}$=$16-bit
        (BF16 or FP16), \texttt{1}$=$32-bit (FP32).
  \item \texttt{tag[1:0]} — protection tier, selecting the bypass
        width~$X$ and hence the partition between Tier-1 (SECDED),
        Tier-2 (SEC), and Tier-3 (bypassed) bit groups within the
        codeword.
\end{itemize}
\noindent
The overhead is 3 bits per 1024 data bits (\textbf{0.29\%} per cacheline),
10--30\(\times\) lower than ARM MTE (3.1\%) or lowRISC (6.25\%). Tags change
only routing/recovery behavior; the physical Partition~A/B boundary
(\S\ref{sec:sram}) is fixed at design time. \husdraft{Tags are set via
overloaded load/store instructions at model load.} If deploying only a single data type, just 2 bits are sufficient.

\subsubsection{Pattern MUX}
\label{sec:pattern_mux}
On writes, the tag drives a Pattern MUX that scatters 128 data bits into Tier-1,
Tier-2, and Tier-3 groups and routes them to codec/partitions; on reads, a
gather restores logical layout. Because valid (data type, tier) combinations are
few and fixed (10--12 for BF16/FP16/FP32), the MUX uses hardwired swizzle
patterns selected by the 3-bit tag, avoiding barrel shifters/crossbars.
\rev{Table~\ref{tab:routing_dse} compares five routing architectures synthesized
in sky130. The hardwired Pattern MUX is ${\sim}4\times$ cheaper than a barrel
shifter or butterfly network and $50\times$ smaller than a full crossbar,
confirming that data type-aware hardwired routing is the right design point for a
fixed set of IEEE~754 formats. The bit-plane interleave itself is a fixed
bit-to-bitline permutation, i.e.\ wiring only: the synthesized interleave router
adds zero logic cells, area, and power, so the sole routing logic is the Pattern
MUX, which is far cheaper than provisioning a multi-bit code
(Table~\ref{tab:mcu-ecc-design-area}).}
\begin{table}[t]
\centering
\footnotesize
\caption{\rev{Routing-architecture design-space exploration (sky130). The
hardwired Pattern MUX is the right design point for a fixed set of data type/tier
swizzle patterns.}}
\label{tab:routing_dse}
\begin{tabular}{lrr}
\toprule
Architecture & Area ($\mu$m$^2$) & Relative \\
\midrule
Pattern MUX (hardwired)  & 2{,}546   & 1.0$\times$ \\
Butterfly                & 9{,}129   & 3.6$\times$ \\
Barrel shifter           & 10{,}090  & 4.0$\times$ \\
Bene\v{s}                & 16{,}816  & 6.6$\times$ \\
Full crossbar            & 127{,}370 & 50$\times$ \\
\bottomrule
\end{tabular}

\end{table}

\subsection{Dual-Partition SRAM}\label{sec:sram}
Requirement~\ref{req:partition} calls for physically separating bypassed
fraction LSBs (Tier~3) from sign/exponent/check bits (Tiers~1/2), so the
non-critical partition can run at lower voltage. This section summarizes
partition sizing, voltage scaling, level-shifter overhead, and macro area.

\subsubsection{Partition Sizing}
\label{sec:partition_sizing}
Each 128-bit logical word is stored as a 136-bit physical word
(128 data + 8 check bits at \(r{=}8\); \S~\ref{sec:check_bit_budget}) and split
across two independently powered partitions:
\begin{itemize}[leftmargin=*]
  \item \textbf{Partition~A} ($V_{\mathrm{nom}}$): \husdraft{holds the
        Tier~1 data bits (sign, exponent, and upper-fraction MSBs)
        and all 8~check bits.}
  \item \textbf{Partition~B} ($V_{\mathrm{low}}$): \husdraft{holds the
        Tier~2 and Tier~3 fraction bits.  Tier~2 bits retain
        detection coverage via the shared Hamming codeword;
        Tier~3 bits are stored without parity.}
\end{itemize}

\noindent
Table~\ref{tab:partition_sizing} shows data type-specific partition widths. Total
physical width is \textbf{always 136 bits}, enabling one SRAM macro across
IEEE~754 formats; only logical routing changes with tag-selected policy.
\husdraft{The invariant width follows from fixing $r{=}8$ across all formats
(\S\ref{sec:check_bit_budget}): regardless of bypass width, the physical word
is always $128 + 8 = 136$~bits.} The A/B boundary is fixed at design time.
\begin{table}[t]
\centering
\footnotesize
\caption{Dual-partition SRAM sizing per 128-bit logical word
         ($r{=}8$).  Partition~A operates at $V_{\mathrm{nom}}$;
         Partition~B at $V_{\mathrm{low}}$.}
\label{tab:partition_sizing}
\begin{tabular}{lccc}
\toprule
 & \textbf{BF16} & \textbf{FP16} & \textbf{FP32} \\
\midrule
Words per codeword $W$
  & 8 & 8 & 4 \\\midrule
Partition~A data (Tier~1)
  & 88\;$(11{\times}8)$ & 96\;$(12{\times}8)$ & 96\;$(24{\times}4)$ \\
Partition~A check bits
  & 8 & 8 & 8 \\\midrule
\textbf{Partition~A total}
  & \textbf{96} & \textbf{104} & \textbf{104} \\\midrule
Partition~B data (Tier~2$+$3)
  & 40\;$(5{\times}8)$ & 32\;$(4{\times}8)$ & 32\;$(8{\times}4)$ \\
\midrule
\textbf{Grand total}
  & \textbf{136} & \textbf{136} & \textbf{136} \\
\bottomrule
\end{tabular}

\end{table}

\subsubsection{Voltage-Scaling Model}\label{sec:voltage_scaling}
\begin{figure}[t]
  \centering
  \includegraphics[width=0.5\linewidth]{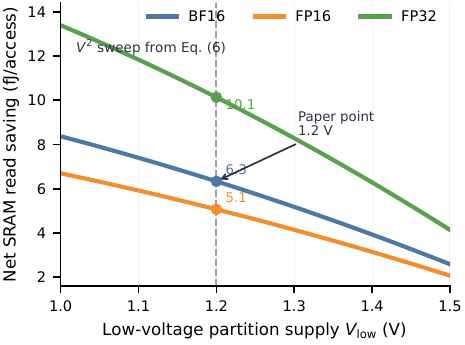}
  \caption{Net SRAM read-energy savings vs.\ $V_{\mathrm{low}}$
           for BF16, FP16, and FP32.  The adopted operating point
           $V_{\mathrm{low}}{=}1.2$\,V (dashed line) yields positive
           net savings for all three formats after level-shifter
           overhead.}
  \label{fig:vlow_sweep}
\end{figure}

Partition~B stores expendable fraction LSBs, so it can tolerate higher raw BER
and operate at reduced voltage (\(V_{\mathrm{low}} < V_{\mathrm{nom}}\)).
\husdraft{Let $f_{\mathrm{nc}}$ denote the fraction of the 136-bit physical
word stored in Partition~B (Table~\ref{tab:partition_sizing}).
For BF16, $f_{\mathrm{nc}} = 40/136 \approx 0.29$;
for FP16 and FP32, $f_{\mathrm{nc}} = 32/136 \approx 0.24$.}
The dynamic and leakage power
savings are:
\begin{align}
  P_{\mathrm{dyn,saved}}  &= f_{\mathrm{nc}}\, P_{\mathrm{dyn}}\,\bigl[1-(V_{\mathrm{low}}/V_{\mathrm{nom}})^{2}\bigr], \label{eq:pdyn} \\
  P_{\mathrm{leak,saved}} &\approx f_{\mathrm{nc}}\, P_{\mathrm{leak}}\,(1-V_{\mathrm{low}}/V_{\mathrm{nom}}). \label{eq:pleak}
\end{align}
\noindent
At target raw BER \(10^{-4}\) (\({\approx}\,0.90\,V_{\min}\)), SRAM-cell
energy drops by about 20\%. We adopt \(V_{\mathrm{low}}=1.2\)\,V
(\(V_{\mathrm{nom}}=1.8\)\,V in sky130), selected from
Figure~\ref{fig:vlow_sweep}.
\rev{Scaled by the Partition-B bypass fraction \(f_{\mathrm{nc}}\) and net of the
integrated level-shifter overhead (Table~\ref{tab:level_shifter}), this is a
\({\sim}17\%\) gross read-energy reduction for BF16 (net positive,
\(+6.3\)\,fJ/access after level shifters) and \({\sim}14\%\) for FP16/FP32
(\(+5.1\) and \(+10.1\)\,fJ/access net).} 

\husdraft{Because Tier~2 bits in Partition~B retain detection coverage
via the shared codeword, corrected-error counts at
$V_{\mathrm{low}}$ provide a built-in feedback signal for
closed-loop voltage tuning, enabling more aggressive voltage
reduction while the correction rate remains below a target
threshold.}

\subsubsection{Level-Shifter Analysis}
\label{sec:level_shifter}
The A/B voltage boundary requires level shifters on all cross-domain signals.
We compare:
\begin{itemize}[leftmargin=*, topsep=0pt]
  \item \textbf{Integrated level shifters} (placed at the partition
        boundary inside the SRAM macro): 0.4\,fJ/transition.
  \item \textbf{Discrete level shifters} (external to the macro):
        5.0\,fJ/transition.
\end{itemize}
\noindent
Integrated shifters are essential: 0.4\,fJ adds only 2--3\% macro-energy
overhead, while 5.0\,fJ discrete shifters consume most voltage-scaling gains.
Voltage is managed at SRAM-bank granularity by a per-bank LDO
(938\,$\mu$m$^2$ per 8-macro bank at 7\,nm). Table~\ref{tab:level_shifter}
shows energy per 136-bit access; integrated shifters keep overhead below 2.1\%
for all formats.
\begin{table}
\footnotesize
\centering
\caption{Level-shifter energy per 136-bit SRAM access.}
\label{tab:level_shifter}
\begin{tabular}{lcccc}
\toprule
 & \textbf{Signals} & \textbf{Energy/sig} & \textbf{Total}
 & \textbf{vs.\ macro} \\
\midrule
Integrated & 40 (BF16) & 0.4\,fJ & 16\,fJ & 2.1\% \\
           & 32 (FP16) & 0.4\,fJ & 12.8\,fJ & 1.7\% \\
           & 32 (FP32) & 0.4\,fJ & 12.8\,fJ & 1.7\% \\
\midrule
Discrete   & 40 (BF16) & 5.0\,fJ & 200\,fJ & 26\% \\
\bottomrule
\end{tabular}

\end{table}

\rev{Beyond per-access energy, integrating the dual-voltage macro imposes
modest physical-layout complexity. Level conversion is required only on the
Partition~B cross-domain signals and is folded into the sense amps and write
drivers rather than a discrete shifter array~\cite{jung2020elssram}, keeping the
energy overhead below 2.1\% (Table~\ref{tab:level_shifter}). The dominant
place-and-route costs are distributing a second ($V_{\mathrm{low}}$) power rail
decoupled from the $V_{\mathrm{nom}}$ periphery and providing n-well/guard-band
separation between the two voltage domains. Because the A/B boundary is fixed at
design time at bank granularity, a single macro serves all formats and only the
per-cacheline tag changes routing; the per-bank LDO switches $V_{\mathrm{low}}$
at model load (${\sim}10$\,ns settle), off the inference critical path. The
resulting macro-area overhead is the ${\sim}4\%$ of Table~\ref{tab:sram_area},
recovered by the energy savings, with full post-layout/silicon validation left
as future work.}

\subsubsection{SRAM Macro Area}
\label{sec:sram_area}
\rev{We synthesize the codec in sky130 and model the SRAM macro in
CACTI~7.0~\cite{cacti}, projecting to 7\,nm (64\,KB, single R/W port) using
published scaling equations~\cite{stillmaker2017} together with a published
7\,nm SRAM bitcell, since CACTI does not directly model sub-22\,nm nodes.} Table~\ref{tab:sram_area} compares monolithic and dual-partition designs.
Dual partitioning incurs \textbf{4\%} area overhead (partition boundary,
integrated shifters, per-bank LDO), recovered by the energy savings in
Figure~\ref{fig:vlow_sweep}.
\begin{table}
\footnotesize
\centering
\caption{CACTI macro area comparison: monolithic vs.\ dual-partition
         SRAM (64\,KB, 7\,nm, single R/W port).}
\label{tab:sram_area}
\begin{tabular}{lcc}
\toprule
Design & Area (mm$^2$) & Relative \\
\midrule
Monolithic 136-bit, $V_{\mathrm{nom}}$ full ECC
  & 0.0281 & 1.00$\times$ \\
Dual-partition 136-bit (A$+$B)
  & 0.0293 & 1.04$\times$ \\
\bottomrule
\end{tabular}

\end{table}

\rev{\textbf{System-level applicability and overheads.} Because UEP is applied
only to inference-produced values (\ref{req:selective}), the result/accumulator
SRAM it modifies is essentially all UEP-eligible by construction, while weights
and inputs remain in separate full-ECC structures. The accumulator is the widest
datapath format ($\geq 2\times$ the operand width) and is always UEP-eligible.
For FP16/BF16 operands with an FP32 accumulator the activation is also wide, so
the bypassable activation-plus-accumulator share of per-MAC traffic is 75\% (the
full-ECC weight is the remaining 25\%). For sub-byte operands the activation has
no safe LSBs (\S~\ref{discussion}); only the wide accumulator is bypassable, and
that share grows from ${\sim}67\%$ (FP8) to ${\sim}80\%$ (FP4) as the operands
shrink. Even in capacity
terms the inference-produced KV-cache reaches ${\sim}52\%$ of weight-plus-KV
storage at a 128K-token context for Llama-3~8B. The corresponding mechanism
overheads are small: the per-cacheline tag is 0.29\% (\S\ref{sec:tag_encoding}),
the full tag\,+\,MUX\,+\,codec datapath is ${<}0.004\%$ of a
$256{\times}256$ systolic tile (Table~\ref{tab:codec_impl}), and the
dual-partition macro's ${\sim}4\%$ area (Table~\ref{tab:sram_area}) is recovered
by the dual-voltage energy saving. These overheads do not erode the savings.}

\rev{\textbf{Read/write latency.} Three effects could matter, all benign. The
tag-driven Pattern MUX adds one pipelined cycle (scatter/gather) with no
throughput loss; it and the UEP codec (smaller than uniform SECDED,
Table~\ref{tab:codec_impl}) sit off the array-read critical path that sets timing.
The dual partition adds no new SRAM---ML accumulator SRAM is already
sub-banked~\cite{amrutur_phd}, so routing Tier-1 vs.\ Tier-2/3 bits to separate
sub-banks is a static wire permutation, zero logic and timing overhead
(\S~\ref{sec:pattern_mux}). Finally, Partition~B at $V_{\mathrm{low}}$ keeps the
clock fixed; unresolved cells appear as the reduced-voltage BER already modeled
(\S~\ref{sec:voltage_scaling}) and, being Tier-2/3 LSBs, fall within
\texttt{Xsafe} with no task-quality effect
(\S~\ref{design-space-exploration-and-reliability}). The only added latency is
the single Pattern-MUX cycle.}

\rev{\textbf{Net benefit.} Taken together, and at iso-reliability for the
critical bits (sign, exponent, and check bits remain at $V_{\mathrm{nom}}$ under
full SECDED, Table~\ref{tab:fit-mttf-summary}), UEP delivers a $27.8\%$ smaller
ECC codec (Table~\ref{tab:codec_impl}) and a ${\sim}17\%$ gross BF16
(${\sim}14\%$ FP16/FP32) non-critical-partition read-energy reduction
(Fig.~\ref{fig:vlow_sweep}; net-positive after integrated level shifters,
Table~\ref{tab:level_shifter})
for a ${\sim}4\%$ macro-area overhead that the energy saving recovers
(Table~\ref{tab:sram_area}) --- without retraining or any change to the stored
numeric representation.}

\noindent
This completes the three-component architecture: the UEP codec
(\S\ref{sec:uep_ecc_codec}) provides graded correction, the per-cacheline tag
(\S\ref{sec:tag}) selects mode per tensor, and the dual-partition SRAM
(\S\ref{sec:sram}) captures energy savings. Section
~\ref{design-space-exploration-and-reliability} evaluates the composed system.

\section{Design Space Exploration and Reliability}\label{design-space-exploration-and-reliability}

\subsection{X-Boundary Sweep and Protection Tier
Comparison}\label{x-boundary-sweep-and-protection-tier-comparison}

\begin{figure}[h]
\centering
\includegraphics[width=0.85\linewidth]{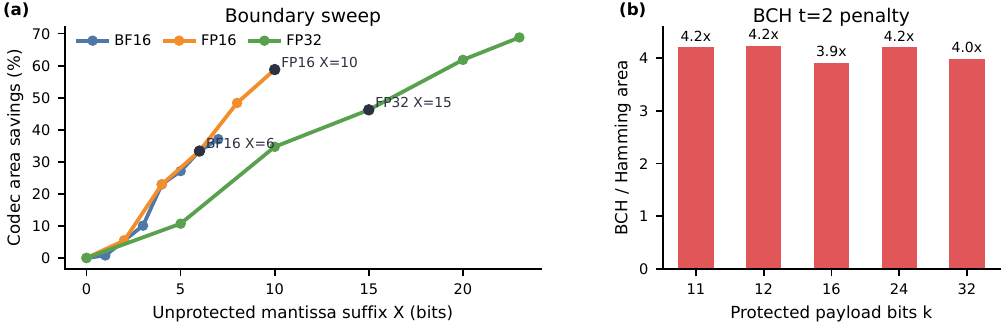}
\caption{\husdraft{Design-space tradeoffs for the \(X\)-boundary sweep.
Left: codec area across BF16, FP16, and FP32 as the protection boundary
\(X\) is varied. Right: relative BCH \(t=2\) and Hamming SECDED area for
the representative protected payload widths used in this section.}}
\label{fig:dse-boundary-bch}
\end{figure}

Given the codec construction in \S\ref{sec:architecture}, this section
evaluates which protection boundary, MCU strategy, and operating point
close the hardware design loop. Sweeping the UEP boundary \(X\) from
full to selective reduces codec area, with the largest gain in BF16 and
smaller but still positive savings in FP32 (see Figure~\ref{fig:dse-boundary-bch} left).
\husdraft{Across the evaluated configurations
(uniform SECDED, selective SECDED at varying \(X\), and our 3-tier UEP)}
and three data types (BF16, FP16, FP32), we use
$k = \text{exponent\_width\_max} + 1$ \husdraft{(the transition-derived
\texttt{exp\_sign} boundary)} as
the representative selective/UEP boundary in this section. A fixed
\(k=8\) is inferior in BF16 (\(k < k_{\text{crit}} = 9\), misses
exponent LSB).

\subsection{MCU-Aware Design Selection}\label{mcu-aware-ecc-design-selection}

BCH $t{=}2$ costs 3.9--4.2$\times$ Hamming
SECDED~\cite{hamming1950,hsiao1970} area across tested codeword widths
(64, 128, 256 bits). Once bit-plane interleaving is included, the
practical area--correction frontier for DNN inference SRAM is dominated
by interleaved Hamming SECDED.
\husdraft{We keep Hamming SECDED as the primary baseline because
lightweight single-symbol correction remains the common starting point
in practical selective or collaborative memory-protection
schemes~\cite{tang2016,lee2022,alam2022}. Table~\ref{tab:mcu-ecc-design-area}
reports the correction/area frontier for the compared options.}

\begin{table}[htbp]
\centering
\footnotesize
\setlength{\tabcolsep}{3pt}
\renewcommand{\arraystretch}{1.0}
\newcolumntype{P}[1]{>{\raggedright\arraybackslash}p{#1\columnwidth}}
\caption{\husdraft{MCU-2 and MCU-3 correction coverage and area relative
to BCH \(t=2\) for the listed Hamming and interleaved protection
options.}}
\label{tab:mcu-ecc-design-area}
\begin{tabular}{P{0.35}P{0.13}P{0.13}P{0.34}}
\toprule
Design & MCU-2 correction & MCU-3 correction & Area (relative to BCH t=2) \\
\midrule
Hamming SECDED (Figure~\ref{fig:dse-boundary-bch}) & 0\% & 0\% & \husdraft{\(\sim\)25\% of BCH \(t{=}2\) (baseline)} \\
Hamming + D=2 interleave & 100\% & 0\% & 56\% of BCH t=2 \\
Hamming + D=3 interleave & 100\% & 100\% & 85\% of BCH t=2 \\
BCH t=2 & 100\% & 52\% & 100\% (3.9--4.2× Hamming) \\
\bottomrule
\end{tabular}

\end{table}

\textbf{Recommendation:} \husdraft{Under the modeled contiguous-upset SRAM
geometry, use Hamming SECDED + D=2 bit-plane interleave as the default
for DNN inference SRAM, and reserve D=3 interleave for
radiation-hard MCU-3 requirements. Table~\ref{tab:mcu-ecc-design-area}
contains the full quantitative comparison.}

\husdraft{The same recommendation also has a simple geometric
interpretation. Without interleaving, one contiguous upset can cross the
mantissa/exponent boundary inside a single protected word; with
bit-plane striping, the same spatial event becomes at most one flipped
bit per word, which is corrected by SECDED.}
\husdraft{Table~\ref{tab:mbu-correction-rates} quantifies this
structure-level effect for the modeled contiguous-upset SRAM pattern.}

\begin{table}[t]
\centering
\footnotesize
\setlength{\tabcolsep}{4pt}
\renewcommand{\arraystretch}{1.0}
\caption{\husdraft{MBU-correction rates for a 4K-word SRAM model under
conventional versus bit-plane SECDED layouts.}}
\label{tab:mbu-correction-rates}
\begin{tabularx}{\linewidth}{@{}C{0.12\linewidth} C{0.22\linewidth} C{0.22\linewidth} Y@{}}
\toprule
\husdraft{MBU} & \husdraft{Conv. SECDED} & \husdraft{Bit-plane SECDED} & \husdraft{Advantage} \\
\midrule
\husdraft{2-bit} & \husdraft{52.41\%} & \husdraft{100\%} & \husdraft{+47.59 pp} \\
\husdraft{3-bit} & \husdraft{28.16\%} & \husdraft{100\%} & \husdraft{+71.84 pp} \\
\husdraft{4-bit} & \husdraft{15.13\%} & \husdraft{100\%} & \husdraft{+84.87 pp} \\
\husdraft{5-bit} & \husdraft{8.84\%} & \husdraft{100\%} & \husdraft{+91.16 pp} \\
\husdraft{6-bit} & \husdraft{4.95\%} & \husdraft{100\%} & \husdraft{+95.05 pp} \\
\husdraft{7-bit} & \husdraft{3.30\%} & \husdraft{100\%} & \husdraft{+96.70 pp} \\
\husdraft{8-bit} & \husdraft{1.77\%} & \husdraft{100\%} & \husdraft{+98.23 pp} \\
\bottomrule
\end{tabularx}

\end{table}

\subsection{Reliability}\label{reliability}

\textbf{FIT rate and mean time to failure (MTTF).} Under the analytic terrestrial-neutron
model, Table~\ref{tab:fit-mttf-summary} reports the FIT/MTTF tradeoff.
All protected configurations remain inside the screening envelope.
Relative to uniform SECDED, selective protection modestly reduces MTTF
because mantissa LSBs are exposed.
\husdraft{When the low-\(V\) partition operates at
BER\({=}10^{-4}\), cross-tier miscorrection in the UEP codec becomes
the dominant failure mode (\(\sim\!127\) FIT, MTTF \(\sim\!900\)
years); the SECDED-protected partition at \(V_{\mathrm{nom}}\) is
unaffected.}

\noindent
\textbf{FIT-level reliability target.} At the assumed
flux~\cite{schroeder2009dramwild,meza2015revisiting,sridharan2015memory,sullivanGpuDram2021},
the dominant undetected multi-bit failure mode in SECDED-protected bits
remains below 0.1 FIT for the \texttt{exp\_sign} operating point.

\begin{table}[t]
\centering
\footnotesize
\setlength{\tabcolsep}{4pt}
\renewcommand{\arraystretch}{1.0}
\caption{\husdraft{FIT/MTTF summary for the 8\,MB SRAM reference design
under no ECC, uniform SECDED, selective SECDED, and 3-tier protection.
The first four rows use the terrestrial-neutron SER model
(1000\,FIT/Mbit)~\cite{baumann2005ser}; the last row accounts for voltage-scaling BER on the
low-\(V\) partition.}}
\label{tab:fit-mttf-summary}
\begin{tabularx}{\linewidth}{@{}L{0.44\linewidth} C{0.20\linewidth} C{0.20\linewidth}@{}}
\toprule
\husdraft{Configuration} & \husdraft{Uncorr. FIT} & \husdraft{MTTF} \\
\midrule
\husdraft{No ECC} & \husdraft{64,000} & \husdraft{1.8 years} \\
\husdraft{Uniform SECDED} & \husdraft{\(\sim0.004\)} & \husdraft{28.5M years} \\
\husdraft{Selective SECDED (BF16 \(X{=}2\))} & \husdraft{\(\sim0.006\)} & \husdraft{19.0M years} \\
\husdraft{3-tier BF16 (ours)} & \husdraft{\(\sim0.005\)} & \husdraft{22.8M years} \\
\husdraft{3-tier BF16 (high BER)\rlap{$^{\dagger}$}} & \husdraft{\(\sim\)127} & \husdraft{\(\sim\)900 years} \\
\bottomrule
\end{tabularx}\\[2pt]
\raggedright\scriptsize
\husdraft{$^{\dagger}$\,Dual-voltage model:
$\lambda_{\mathrm{SEU}}{=}10^{-12}$/bit/hr (Partition~A),
BER${=}10^{-4}$ (Partition~B); cross-tier miscorrection dominated.}

\end{table}

\rev{The ${\sim}900$-year row is the most aggressive operating point and
reflects only Partition~B: sign, exponent, and all check bits remain at
$V_{\mathrm{nom}}$ under full SECDED, so critical-bit MTTF is unchanged
(millions of years), and a cross-tier miscorrection lands in a Tier-2 fraction
LSB that \S\ref{bit-position-sensitivity-results} shows is fault-tolerant---the
end-to-end results (Table~\ref{tab:multimodel-tiered-ber}) confirm task quality
holds at this point. The operating point is tunable via the Tier-2 detection
feedback: a $10\times$ lower Partition-B BER raises this MTTF to ${\sim}9{,}000$
years and $100\times$ to ${\sim}89{,}000$ years. Even the aggressive
${\sim}900$-year point exceeds a 3--5 year accelerator service life by
${\sim}180$--$300\times$, so the residual events are rare, benign LSB
miscorrections rather than critical data corruption.}

\subsection{Validation}\label{validation}

\husdraft{\textbf{System-level validation (Llama-3 8B).}
Table~\ref{tab:llama3-tiered-ber} reports BF16 perplexity and tier-1
miscorrection counts across representative
\((\mathrm{BER}_p,\mathrm{BER}_u)\) operating points, and serves as the
task-level check for the selective-exposure assumptions in the
voltage-scaling and selective-protection analysis.}

\begin{table}[t]
\centering
\scriptsize
\setlength{\tabcolsep}{2pt}
\renewcommand{\arraystretch}{1.04}
\caption{\husdraft{Llama-3 8B WikiText-2 perplexity under 3-tier ECC
(baseline PPL = 5.3659). BER\(_p\) and BER\(_u\) denote protected-tier
and unprotected-tier bit error rates; the table reports PPL change and
tier-1 miscorrection counts for each operating point.}}
\label{tab:llama3-tiered-ber}
\begin{tabularx}{\linewidth}{@{}L{0.24\linewidth} C{0.09\linewidth} C{0.10\linewidth} C{0.11\linewidth} C{0.11\linewidth} Y@{}}
\toprule
\husdraft{Scenario} & \husdraft{BER\(_p\)} & \husdraft{BER\(_u\)} & \husdraft{PPL (BF16)} & \husdraft{\(\Delta\)\%} & \husdraft{T1 Miscorr.} \\
\midrule
\husdraft{Ultra-conservative} & \husdraft{0} & \husdraft{\(10^{-5}\)} & \husdraft{5.3646} & \husdraft{-0.02} & \husdraft{---} \\
\husdraft{Moderate} & \husdraft{0} & \husdraft{\(10^{-3}\)} & \husdraft{5.3634} & \husdraft{-0.05} & \husdraft{---} \\
\husdraft{Aggressive} & \husdraft{0} & \husdraft{\(10^{-2}\)} & \husdraft{5.3674} & \husdraft{+0.03} & \husdraft{---} \\
\husdraft{Realistic-moderate} & \husdraft{\(10^{-7}\)} & \husdraft{\(10^{-3}\)} & \husdraft{5.3621} & \husdraft{-0.07} & \husdraft{0 / 18{,}181} \\
\husdraft{Stressed-protected} & \husdraft{\(10^{-6}\)} & \husdraft{\(10^{-2}\)} & \husdraft{5.3682} & \husdraft{+0.04} & \husdraft{1 / 145{,}748} \\
\bottomrule
\end{tabularx}

\end{table}

\rev{To show the selective-protection boundary generalizes beyond Llama-3~8B,
we run the same tiered-BER end-to-end protocol on one model per category
(Table~\ref{tab:multimodel-tiered-ber}). At the realistic operating point all
categories preserve task quality; the five non-diffusion models additionally
hold at the Stressed-protected point. Diffusion (SD3.5-Medium) holds at the
realistic protected-tier rate ($\mathrm{BER}_p{=}10^{-7}$) and degrades only
beyond the operating regime, consistent with it being the most fault-sensitive
class that sets the conservative floor.}

\begin{table}[t]
\centering
\scriptsize
\setlength{\tabcolsep}{3pt}
\renewcommand{\arraystretch}{1.0}
\caption{\rev{Tiered-BER end-to-end validation extended to one model per
category (BF16) at the Realistic-moderate operating point
($\mathrm{BER}_p{=}10^{-7}$, $\mathrm{BER}_u{=}10^{-3}$). Task quality is
preserved across all categories. The five non-diffusion models also hold at the
Stressed-protected point ($\mathrm{BER}_p{=}10^{-6}$); diffusion, the
most fault-sensitive class that sets the conservative floor, degrades only
beyond that regime. For SD3.5-Medium the Baseline column is the
near-clean reproducibility reference (${\sim}33$\,dB, ultra-conservative point),
as a clean-vs-clean PSNR is undefined; at realistic-moderate, PSNR
$29.83$\,dB and SSIM $0.95$ (image quality preserved).}}
\label{tab:multimodel-tiered-ber}
\begin{tabularx}{\linewidth}{@{}L{0.30\linewidth} L{0.24\linewidth} C{0.18\linewidth} C{0.18\linewidth}@{}}
\toprule
Model & Metric & Baseline & Realistic-mod. \\
\midrule
Mistral-7B    & PPL ($\downarrow$)      & 5.4415 & 5.4431 \\
ViT-L/16      & Top-1 \%                & 80.30  & 80.35 \\
BERT-base     & SST-2 Acc \%            & 92.43  & 92.32 \\
ResNet-50     & Top-1 \%                & 81.25  & 81.25 \\
Whisper-small & WER \% ($\downarrow$)   & 17.39  & 17.48 \\
SD3.5-Medium  & PSNR (dB)               & ${\sim}33$\,(ref) & 29.83 \\
\bottomrule
\end{tabularx}

\end{table}


\paragraph{End-to-end validation chain}
\husdraft{The chain has four steps: (1) RTL fixes which bits are exposed,
(2) the Stutz mapping sets BER at the adopted \(V_\text{low}\)
operating point~\cite{stutz2021}, (3) BER-matched Ares heatmaps provide
a conservative all-bit reference, and (4) direct selective BF16 runs
confirm the milder LSB-only regime.
The per-data type RTL operating points, following the \textit{exp\_sign}
partitioning of \S~\ref{sec:uep_ecc_codec} (a conservative boundary distinct
from the \texttt{Xsafe} floors), are
\(k{=}11\) for BF16 (5 unprotected LSBs, bits 0--4),
\(k{=}12\) for FP16 (4 LSBs, bits 0--3), and
\(k{=}24\) for FP32 (8 LSBs, bits 0--7);
shared BER and read/write-energy assumptions follow the dual-voltage
model in \S~\ref{sec:sram}.
Table~\ref{tab:llama3-tiered-ber} provides the task-metric check.}


\section{Discussion, Limitations, and Future Work}\label{discussion}\label{limitations-and-future-work}

\looseness=-1
\rev{UEP is applicable to the wide floating-point formats in which results are
produced and accumulated---the FP16/BF16/FP32 partial sums and accumulators of the
matmul/result path. These wide formats carry the measured \texttt{Xsafe} headroom
(6/4/15) and therefore expose mantissa LSBs that selective ECC can safely leave
unprotected. UEP does \emph{not} apply to narrow stored operands: as weights
\emph{and} activations move to sub-byte formats such as FP8/FP4 (e.g.\ E4M3/E2M1),
they have too few mantissa bits to admit a safe unprotected region, so UEP yields no
reduction on narrow-format storage and such operands retain full protection.
Crucially, this boundary does not shrink UEP's relevance---it grows it: as the 
same on-chip bandwidth is spent on ever-narrower operands, each transfer brings 
a larger \emph{number} of elements; accumulating more elements per operation only
increases the need for the wide, higher-precision accumulation formats (FP16/BF16/FP32)
where UEP applies. Moreover, as operands narrow, the accumulation/result path stays 
wide (FP16/BF16/FP32) while the operands' byte footprint shrinks, so the wide, 
UEP-eligible accumulator becomes a growing share of on-chip floating-point traffic
in absolute bytes---the relevant figure of merit for an ECC budget---expanding rather
than eroding the opportunity for area and energy savings. SERA-Float~\cite{mishra2025}
remains complementary rather than competing: it changes the stored representation 
lossily, whereas our UEP policy keeps the representation fixed and varies ECC strength
by bit significance, so the two can compose by compressing first and then applying 
tiered protection to the retained bits.\label{sera-float-comparison}}

\looseness=-1
\husdraft{Three limitations remain: synthetic plus bounded-neighborhood
injection only partially captures field-correlated fault processes~\cite{sullivanGpuDram2021,meza2015revisiting,sridharan2015memory,alam2022},
the study does not yet include a full accelerator prototype with the
selective-ECC controller in the execution loop, and dual-voltage SRAM
integration is constrained by bank-level partition granularity,
rail-switch latency, and physical-routing overhead. The next steps are
clear: incorporate trace-derived spatial/temporal fault logs, perform
post-layout or silicon-proximate validation of the full selective-ECC
stack, and extend deployment from RF/controller-local opportunities to
mixed-type L1/L2 cache hierarchies with per-cacheline data type metadata.}

\looseness=-1
\rev{The decision to keep weights and inputs under full protection is not
merely conservative: because they are loaded once and reused throughout
inference, a single persistent fault propagates to every downstream use---a
sign-bit flip in BF16 weights raises Llama-3~8B perplexity by over four orders
of magnitude, to ${\sim}1.3{\times}10^{5}$. Yet weight mantissa LSBs are
themselves tolerant, so a weight-side UEP whose protection cost is amortized
over the load-once lifetime is a promising direction for future work. This need
only grows as models are compressed: quantization or pruning removes redundant
bits, so each retained weight bit becomes more critical, and a
significance-aware, per-bit-position scheme is exactly the ``more intelligent''
error correction such compressed weights would require.}

\section{Conclusion}\label{conclusion}

\looseness=-1
\husdraft{This paper establishes that the mantissa-to-exponent sensitivity
transition is a data type-level IEEE~754 property, not a model artifact, and
that conservative floors (FP16 \texttt{Xsafe}=6, BF16 \texttt{Xsafe}=4,
FP32 \texttt{Xsafe}=15) remain stable across the evaluated
workloads. Translating this boundary into data type-aware UEP delivers
efficiency gains without retraining: a 37.5\% conservative FP16
reduction floor, 56.25--62.5\% workload-aware FP16 reduction, \rev{a 27.8\% ECC
codec area reduction, and \({\sim}17\%\) gross BF16 read-energy reduction from
dual-voltage operation (the dual-partition macro adds \({\sim}4\%\) area,
recovered by these savings)}. Reliability stays within the
same hierarchy-local envelope: MCU-3 results preserve the boundary, and
\husdraft{FIT-level modeling stays below 10~FIT under the terrestrial-SER
regime (MTTF above 19.0M years); under dual-voltage operation
(BER\({=}10^{-4}\)), cross-tier miscorrection dominates at
\(\sim\!\)127~FIT (MTTF \(\sim\!\)900 years).} Overall, data type-aware ECC is not a secondary
optimization; it is a practical memory co-design lever for ML
accelerators, especially on on-chip SRAM/RF paths where data format is
explicit and policy can align directly with floating-point semantics.}





\bibliographystyle{ACM-Reference-Format}
\bibliography{references}

@manual{ieee754_2019,
  title        = {IEEE Standard for Floating-Point Arithmetic},
  author       = {{IEEE Computer Society}},
  organization = {Institute of Electrical and Electronics Engineers},
  address      = {New York, NY, USA},
  year         = {2019},
  month        = jul,
  note         = {IEEE Std 754-2019 (Revision of IEEE 754-2008)},
  doi          = {10.1109/IEEESTD.2019.8766229},
  url          = {https://ieeexplore.ieee.org}
}

@inproceedings{frugalEcc,
  author = {Kim, Jungrae and Sullivan, Michael B. and Gong, Seong-Lyong and Erez, Mattan},
  title = {Frugal {ECC}: Efficient and Versatile Memory Error Protection Through Fine-Grained Compression},
  booktitle = {{SC}},
  year = {2015},
  address = {Austin, TX, USA},
  pages = {12:1--12:12},
  numpages = {12},
  publisher = {ACM},
  doi = {10.1145/2807591.2807659}
}

@article{tang2016,
  author = {Tang, Hoyoung and Park, Jongsun},
  title = {Unequal-Error-Protection Error Correction Codes for the Embedded Memories in Digital Signal Processors},
  journal = {{IEEE} Transactions on Very Large Scale Integration ({VLSI}) Systems},
  volume = {24},
  number = {6},
  pages = {2397--2401},
  year = {2016},
  publisher = {IEEE},
  doi = {10.1109/TVLSI.2015.2497368}
}

@inproceedings{stutz2021,
  author = {Stutz, David and Chandramoorthy, Nandhini and Hein, Matthias and Schiele, Bernt},
  title = {Bit Error Robustness for Energy-Efficient {DNN} Accelerators},
  booktitle = {Proceedings of {MLSys}},
  year = {2021},
  address = {Virtual},
  pages = {1--21},
  numpages = {21},
  publisher = {MLSys},
  url = {https://proceedings.mlsys.org/paper_files/paper/2021/file/2f9b1b6b29361118f630783c19891ea0-Paper.pdf}
}

@inproceedings{koppula2019,
  author = {Koppula, Skanda and Orosa, Lois and Ya\u{g}l{\i}k\c{c}{\i}, A. Giray and Azizi, Roknoddin and Shahroodi, Taha and Kanellopoulos, Konstantinos and Mutlu, Onur},
  title = {{EDEN}: Enabling Energy-Efficient, High-Performance Deep Neural Network Inference Using Approximate {DRAM}},
  booktitle = {{MICRO}},
  year = {2019},
  address = {Columbus, OH, USA},
  pages = {218--231},
  numpages = {14},
  publisher = {ACM},
  doi = {10.1145/3352460.3358280}
}

@inproceedings{zheng2017,
  author = {Zheng, Ruohuang and Huang, Michael C.},
  title = {Redundant Memory Array Architecture for Efficient Selective Protection},
  booktitle = {{ISCA}},
  year = {2017},
  address = {Toronto, ON, Canada},
  pages = {214--227},
  numpages = {14},
  publisher = {ACM},
  doi = {10.1145/3079856.3080213}
}

@inproceedings{mishra2025,
  author = {Mishra, Vishesh and Traiola, Marcello and Kritikakou, Angeliki and Sentieys, Olivier and Chatterjee, Urbi},
  title = {{SERA}-Float: A Soft Error Resilient Approximate Floating-Point Computing Format},
  booktitle = {{ICCAD}},
  year = {2025},
  address = {Munich, Germany},
  pages = {1--9},
  numpages = {9},
  publisher = {IEEE},
  url = {https://inria.hal.science/hal-05333255v1/file/SERA_FLOAT_v2.pdf}
}

@article{catalan2025,
  author = {Catal{\'a}n, Izan and Flich, Jos{\'e} and Hern{\'a}ndez, Carles},
  title = {Exploiting Neural Networks Bit-Level Redundancy to Mitigate the Impact of Faults at Inference},
  journal = {The Journal of Supercomputing},
  volume = {81},
  pages = {183},
  year = {2025},
  publisher = {Springer},
  doi = {10.1007/s11227-024-06693-7}
}

@inproceedings{qureshi2015,
  author = {Qureshi, Moinuddin K. and Kim, Dae-Hyun and Khan, Samira Manabi and Nair, Prashant J. and Mutlu, Onur},
  title = {{AVATAR}: A Variable-Retention-Time ({VRT}) Aware Refresh for {DRAM} Systems},
  booktitle = {{DSN}},
  year = {2015},
  address = {Rio de Janeiro, Brazil},
  pages = {427--437},
  numpages = {11},
  publisher = {IEEE},
  doi = {10.1109/DSN.2015.58}
}

@inproceedings{mathew2018,
  author = {Mathew, Deepak M. and Schultheis, Martin and Rheinl{\"a}nder, Carl C. and Sudarshan, Chirag and Weis, Christian and Wehn, Norbert and Jung, Matthias},
  title = {An Analysis on Retention Error Behavior and Power Consumption of Recent {DDR4} {DRAM}s},
  booktitle = {{DATE}},
  year = {2018},
  address = {Dresden, Germany},
  pages = {293--296},
  numpages = {4},
  publisher = {IEEE},
  doi = {10.23919/DATE.2018.8342023}
}

@inproceedings{oh2016,
  author = {Oh, Byoungchan and Abeyratne, Nilmini and Ahn, Jeongseob and Dreslinski, Ronald G. and Mudge, Trevor},
  title = {Enhancing {DRAM} Self-Refresh for Idle Power Reduction},
  booktitle = {{ISLPED}},
  year = {2016},
  address = {San Francisco, CA, USA},
  pages = {254--259},
  numpages = {6},
  publisher = {ACM},
  doi = {10.1145/2934583.2934632}
}

@inproceedings{approxMem2018,
  author = {Nguyen, Duy Thanh and Kim, Hyun and Lee, Hyuk-Jae and Chang, Ik Joon},
  title = {An Approximate Memory Architecture for a Reduction of Refresh Power Consumption in Deep Learning Applications},
  booktitle = {{ISCAS}},
  year = {2018},
  address = {Florence, Italy},
  pages = {1--5},
  numpages = {5},
  publisher = {IEEE},
  doi = {10.1109/ISCAS.2018.8351021}
}

@inproceedings{aftEcc2023,
  author = {Sullivan, Michael B. and Ibn Ziad, Mohamed Tarek and Jaleel, Aamer and Keckler, Stephen W.},
  title = {Implicit Memory Tagging: No-Overhead Memory Safety Using Alias-Free Tagged {ECC}},
  booktitle = {{ISCA}},
  year = {2023},
  address = {Orlando, FL, USA},
  pages = {1--14},
  numpages = {14},
  publisher = {ACM},
  doi = {10.1145/3579371.3589079}
}

@inproceedings{palframan2014,
  author = {Palframan, David J. and Kim, Nam Sung and Lipasti, Mikko H.},
  title = {Precision-Aware Soft Error Protection for {GPU}s},
  booktitle = {{HPCA}},
  year = {2014},
  address = {Orlando, FL, USA},
  pages = {49--59},
  numpages = {11},
  publisher = {IEEE},
  doi = {10.1109/HPCA.2014.6835966}
}

@inproceedings{sullivanGpuDram2021,
  author = {Sullivan, Michael B. and Saxena, Nirmal and O'Connor, Mike and Lee, Donghyuk and Racunas, Paul and Hukerikar, Saurabh and Tsai, Timothy and Hari, Siva and Keckler, Stephen W.},
  title = {Characterizing and Mitigating Soft Errors in {GPU} {DRAM}},
  booktitle = {{MICRO}},
  year = {2021},
  publisher = {ACM},
  doi = {10.1145/3466752.3480111}
}

@article{cacti,
  author = {Balasubramonian, Rajeev and Kahng, Andrew B. and Muralimanohar, Naveen and Shafiee, Ali and Srinivas, Vaishnav},
  title = {{CACTI} 7: New Tools for Interconnect Exploration in Innovative Off-Chip Memories},
  journal = {ACM Trans. Archit. Code Optim.},
  volume = {14},
  number = {2},
  pages = {14:1--14:25},
  year = {2017},
  publisher = {ACM},
  doi = {10.1145/3085572}
}

@inproceedings{reagen2018,
  author = {Reagen, Brandon and Gupta, Udit and Pentecost, Lillian and Whatmough, Paul and Lee, Sae Kyu and Mulholland, Niamh and Brooks, David and Wei, Gu-Yeon},
  title = {Ares: A Framework for Quantifying the Resilience of Deep Neural Networks},
  booktitle = {{DAC}},
  year = {2018},
  address = {San Francisco, CA, USA},
  pages = {17:1--17:6},
  numpages = {6},
  publisher = {ACM},
  doi = {10.1145/3195970.3195997}
}

@inproceedings{zhang2018,
  author = {Zhang, Jeff and Rangineni, Kartheek and Ghodsi, Zahra and Garg, Siddharth},
  title = {{ThUnderVolt}: Enabling Aggressive Voltage Underscaling and Timing Error Resilience for Energy Efficient Deep Learning Accelerators},
  booktitle = {{DAC}},
  year = {2018},
  address = {San Francisco, CA, USA},
  pages = {19:1--19:6},
  numpages = {6},
  publisher = {ACM},
  doi = {10.1145/3195970.3196129}
}

@inproceedings{lee2022,
  author = {Lee, Young Seo and Koo, Gunjae and Gong, Young-Ho and Chung, Sung Woo},
  title = {Stealth {ECC}: A Data-Width Aware Adaptive {ECC} Scheme for {DRAM} Error Resilience},
  booktitle = {{DATE}},
  year = {2022},
  address = {Antwerp, Belgium},
  pages = {382--387},
  numpages = {6},
  publisher = {IEEE},
  doi = {10.23919/DATE54114.2022.9774775}
}

@inproceedings{alam2022,
  author = {Alam, Irina and Gupta, Puneet},
  title = {{COMET}: On-Die and In-Controller Collaborative Memory {ECC} Technique for Safer and Stronger Correction of {DRAM} Errors},
  booktitle = {{DSN}},
  year = {2022},
  address = {Baltimore, MD, USA},
  pages = {124--136},
  numpages = {13},
  publisher = {IEEE},
  doi = {10.1109/DSN53405.2022.00024},
  url = {https://ieeexplore.ieee.org/document/9833662}
}

@misc{xie2025,
  author = {Xie, Rui and Fang, Yunhua and Haq, Asad Ul and Ma, Linsen and Sen, Sanchari and Venkataramani, Swagath and Liu, Liu and Zhang, Tong},
  title = {Making Strong Error-Correcting Codes Work Effectively for {HBM} in {AI} Inference},
  year = {2025},
  eprint = {2512.18152},
  eprinttype = {arXiv},
  doi = {10.48550/arXiv.2512.18152},
  url = {https://arxiv.org/abs/2512.18152}
}

@misc{xie2025bitcost,
  author = {Xie, Rui and Haq, Asad Ul and Fang, Yunhua and Ma, Linsen and Sen, Sanchari and Venkataramani, Swagath and Liu, Liu and Zhang, Tong},
  title = {Breaking the {HBM} Bit Cost Barrier: Domain-Specific {ECC} for {AI} Inference Infrastructure},
  year = {2025},
  eprint = {2507.02654},
  eprinttype = {arXiv},
  doi = {10.48550/arXiv.2507.02654},
  url = {https://arxiv.org/abs/2507.02654}
}

@inproceedings{ha2025,
  author = {Ha, Taeuk and Kim, Gyuri and Kim, Chanki and Kim, Sang-Hyo},
  title = {Chipkill-Level {ECC} Using 4-Bit Symbol {Reed--Solomon} Codes for {DDR5} {DRAM}},
  booktitle = {{ICTC}},
  year = {2025},
  month = oct,
  address = {Jeju, South Korea},
  pages = {371--375},
  numpages = {5},
  publisher = {IEEE},
  isbn = {979-8-3315-5678-5},
  doi = {10.1109/ICTC66702.2025.11387968},
  url = {https://ieeexplore.ieee.org/document/11387968}
}

@inproceedings{wang2020,
  author = {Wang, Yaohua and Orosa, Lois and Peng, Xiangjun and Guo, Yang and Ghose, Saugata and Patel, Minesh and Kim, Jeremie S. and G{\'o}mez Luna, Juan and Sadrosadati, Mohammad and Ghiasi, Nika Mansouri and Mutlu, Onur},
  title = {{FIGARO}: Improving System Performance via Fine-Grained In-{DRAM} Data Relocation and Caching},
  booktitle = {{MICRO}},
  year = {2020},
  address = {Virtual (originally Athens, Greece)},
  pages = {235--249},
  numpages = {15},
  publisher = {IEEE},
  url = {https://ieeexplore.ieee.org/document/9251865}
}

@inproceedings{devlin2019bert,
  author = {Devlin, Jacob and Chang, Ming-Wei and Lee, Kenton and Toutanova, Kristina},
  title = {{BERT}: Pre-training of Deep Bidirectional Transformers for Language Understanding},
  booktitle = {{NAACL-HLT}},
  year = {2019},
  url = {https://arxiv.org/abs/1810.04805}
}

@inproceedings{dosovitskiy2021vit,
  author = {Dosovitskiy, Alexey and Beyer, Lucas and Kolesnikov, Alexander and Weissenborn, Dirk and Zhai, Xiaohua and Unterthiner, Thomas and Dehghani, Mostafa and Minderer, Matthias and Heigold, Georg and Gelly, Sylvain and Uszkoreit, Jakob and Houlsby, Neil},
  title = {An Image is Worth 16x16 Words: Transformers for Image Recognition at Scale},
  booktitle = {{ICLR}},
  year = {2021},
  url = {https://arxiv.org/abs/2010.11929}
}

@inproceedings{liu2021swin,
  author = {Liu, Ze and Lin, Yutong and Cao, Yue and Hu, Han and Wei, Yixuan and Zhang, Zheng and Lin, Stephen and Guo, Baining},
  title = {Swin Transformer: Hierarchical Vision Transformer using Shifted Windows},
  booktitle = {{ICCV}},
  year = {2021},
  url = {https://arxiv.org/abs/2103.14030}
}

@misc{radford2022whisper,
  author = {Radford, Alec and Kim, Jong Wook and Xu, Tao and Brockman, Greg and McLeavey, Christine and Sutskever, Ilya},
  title = {Robust Speech Recognition via Large-Scale Weak Supervision},
  year = {2022},
  eprint = {2212.04356},
  eprinttype = {arXiv},
  url = {https://arxiv.org/abs/2212.04356}
}

@inproceedings{he2016resnet,
  author = {He, Kaiming and Zhang, Xiangyu and Ren, Shaoqing and Sun, Jian},
  title = {Deep Residual Learning for Image Recognition},
  booktitle = {{CVPR}},
  year = {2016},
  url = {http://openaccess.thecvf.com/content_cvpr_2016/html/He_Deep_Residual_Learning_CVPR_2016_paper.html}
}

@inproceedings{tan2019efficientnet,
  author = {Tan, Mingxing and Le, Quoc V.},
  title = {{EfficientNet}: Rethinking Model Scaling for Convolutional Neural Networks},
  booktitle = {{ICML}},
  year = {2019},
  url = {https://research.google/pubs/efficientnet-rethinking-model-scaling-for-convolutional-neural-networks/}
}

@inproceedings{howard2019mobilenetv3,
  author = {Howard, Andrew and Sandler, Mark and Chu, Grace and Chen, Liang-Chieh and Chen, Bo and Tan, Mingxing and Wang, Weijun and Zhu, Yukun and Pang, Ruoming and Vasudevan, Vijay and Le, Quoc V. and Adam, Hartwig},
  title = {Searching for {MobileNetV3}},
  booktitle = {{ICCV}},
  year = {2019},
  url = {http://openaccess.thecvf.com/content_ICCV_2019/html/Howard_Searching_for_MobileNetV3_ICCV_2019_paper.html}
}

@inproceedings{peebles2023dit,
  author = {Peebles, William and Xie, Saining},
  title = {Scalable Diffusion Models with Transformers},
  booktitle = {{ICCV}},
  year = {2023},
  url = {https://www.wpeebles.com/DiT.html}
}

@inproceedings{rombach2022ldm,
  author = {Rombach, Robin and Blattmann, Andreas and Lorenz, Dominik and Esser, Patrick and Ommer, Bj{\"o}rn},
  title = {High-Resolution Image Synthesis with Latent Diffusion Models},
  booktitle = {{CVPR}},
  year = {2022},
  url = {https://arxiv.org/abs/2112.10752}
}

@inproceedings{socher2013sst,
  author = {Socher, Richard and Perelygin, Alex and Wu, Jean and Chuang, Jason and Manning, Christopher D. and Ng, Andrew and Potts, Christopher},
  title = {Recursive Deep Models for Semantic Compositionality Over a Sentiment Treebank},
  booktitle = {{EMNLP}},
  year = {2013},
  pages = {1631--1642},
  publisher = {Association for Computational Linguistics},
  url = {https://aclanthology.org/D13-1170/}
}

@inproceedings{panayotov2015librispeech,
  author = {Panayotov, Vassil and Chen, Guoguo and Povey, Daniel and Khudanpur, Sanjeev},
  title = {{LibriSpeech}: An {ASR} Corpus Based on Public Domain Audio Books},
  booktitle = {{ICASSP}},
  year = {2015},
  url = {https://us.openslr.org/resources/12/about.html}
}

@article{russakovsky2015imagenet,
  author = {Russakovsky, Olga and Deng, Jia and Su, Hao and Krause, Jonathan and Satheesh, Sanjeev and Ma, Sean and Huang, Zhiheng and Karpathy, Andrej and Khosla, Aditya and Bernstein, Michael and Berg, Alexander C. and Fei-Fei, Li},
  title = {{ImageNet} Large Scale Visual Recognition Challenge},
  journal = {International Journal of Computer Vision},
  volume = {115},
  number = {3},
  pages = {211--252},
  year = {2015},
  publisher = {Springer},
  doi = {10.1007/s11263-015-0816-y},
  url = {https://image-net.org/challenges/LSVRC/2015/}
}

@inproceedings{zellers2019hellaswag,
  author = {Zellers, Rowan and Holtzman, Ari and Bisk, Yonatan and Farhadi, Ali and Choi, Yejin},
  title = {{HellaSwag}: Can a Machine Really Finish Your Sentence?},
  booktitle = {Proceedings of the 57th Annual Meeting of the Association for Computational Linguistics},
  year = {2019},
  pages = {4791--4800},
  publisher = {Association for Computational Linguistics},
  doi = {10.18653/v1/P19-1472},
  url = {https://aclanthology.org/P19-1472/}
}

@inproceedings{dodge2021c4,
  author = {Dodge, Jesse and Sap, Maarten and Marasovi{\'c}, Ana and Agnew, William and Ilharco, Gabriel and Groeneveld, Dirk and Mitchell, Margaret and Gardner, Matt},
  title = {Documenting Large Webtext Corpora: A Case Study on the Colossal Clean Crawled Corpus},
  booktitle = {Proceedings of the 2021 Conference on Empirical Methods in Natural Language Processing},
  year = {2021},
  pages = {1286--1305},
  publisher = {Association for Computational Linguistics},
  doi = {10.18653/v1/2021.emnlp-main.98},
  url = {https://anthology.aclweb.org/2021.emnlp-main.98/}
}

@inproceedings{merity2017pointer,
  author = {Merity, Stephen and Xiong, Caiming and Bradbury, James and Socher, Richard},
  title = {Pointer Sentinel Mixture Models},
  booktitle = {{ICLR}},
  year = {2017},
  url = {https://openreview.net/forum?id=Byj72udxe&noteId=Byj72udxe}
}

@article{wilcoxon1945,
  author = {Wilcoxon, Frank},
  title = {Individual Comparisons by Ranking Methods},
  journal = {Biometrics Bulletin},
  volume = {1},
  number = {6},
  pages = {80--83},
  year = {1945},
  url = {https://sci2s.ugr.es/keel/pdf/algorithm/articulo/wilcoxon1945.pdf}
}

@article{clopper1934,
  author = {Clopper, C. J. and Pearson, E. S.},
  title = {The Use of Confidence or Fiducial Limits Illustrated in the Case of the Binomial},
  journal = {Biometrika},
  volume = {26},
  number = {4},
  pages = {404--413},
  year = {1934},
  url = {https://barestatistics.nl/uploads/1/1/7/9/11797954/clopper__pearson_1934.pdf}
}

@article{mcnemar1947,
  author = {McNemar, Quinn},
  title = {Note on the Sampling Error of the Difference Between Correlated Proportions or Percentages},
  journal = {Psychometrika},
  volume = {12},
  number = {2},
  pages = {153--157},
  year = {1947},
  doi = {10.1007/BF02295996},
  url = {https://link.springer.com/content/pdf/10.1007/BF02295996.pdf}
}

@inproceedings{zhang2018lpips,
  author = {Zhang, Richard and Isola, Phillip and Efros, Alexei A. and Shechtman, Eli and Wang, Oliver},
  title = {The Unreasonable Effectiveness of Deep Features as a Perceptual Metric},
  booktitle = {{CVPR}},
  year = {2018},
  url = {http://richzhang.github.io/PerceptualSimilarity/}
}

@article{joshi2020pcm,
  author = {Joshi, Vinay and Le Gallo, Manuel and Haefeli, Simon and Boybat, Irem and Nandakumar, S. R. and Piveteau, Christophe and Dazzi, Martino and Rajendran, Bipin and Sebastian, Abu and Eleftheriou, Evangelos},
  title = {Accurate Deep Neural Network Inference Using Computational Phase-Change Memory},
  journal = {Nature Communications},
  volume = {11},
  pages = {2473},
  year = {2020},
  publisher = {Nature},
  doi = {10.1038/s41467-020-16108-9},
  url = {https://arxiv.org/abs/1906.03138}
}

@article{narayanan2017nvm,
  author = {Narayanan, Pritish and Fumarola, Alessandro and Sanches, Lucas L. and Hosokawa, Kohji and Lewis, Scott C. and Shelby, Robert M. and Burr, Geoffrey W.},
  title = {Toward On-Chip Acceleration of the Backpropagation Algorithm Using Nonvolatile Memory},
  journal = {IBM Journal of Research and Development},
  volume = {61},
  number = {4/5},
  year = {2017},
  url = {https://research.ibm.com/publications/toward-on-chip-acceleration-of-the-backpropagation-algorithm-using-nonvolatile-memory}
}

@article{goldberg1991,
  author = {Goldberg, David},
  title = {What Every Computer Scientist Should Know About Floating-Point Arithmetic},
  journal = {ACM Computing Surveys},
  volume = {23},
  number = {1},
  pages = {5--48},
  year = {1991},
  doi = {10.1145/103162.103163},
  url = {https://dl.acm.org/doi/10.1145/103162.103163}
}

@article{hamming1950,
  author = {Hamming, Richard W.},
  title = {Error Detecting and Error Correcting Codes},
  journal = {Bell System Technical Journal},
  volume = {29},
  number = {2},
  pages = {147--160},
  year = {1950},
  doi = {10.1002/j.1538-7305.1950.tb00463.x},
  url = {https://ui.adsabs.harvard.edu/abs/1950BSTJ...29..147H/abstract}
}

@article{hsiao1970,
  author = {Hsiao, M. Y.},
  title = {A Class of Optimal Minimum Odd-Weight-Column {SEC-DED} Codes},
  journal = {IBM Journal of Research and Development},
  volume = {14},
  number = {4},
  pages = {395--401},
  year = {1970},
  doi = {10.1147/rd.144.0395},
  url = {https://ieeexplore.ieee.org/document/5391627}
}

@inproceedings{schroeder2009dramwild,
  author = {Schroeder, Bianca and Pinheiro, Eduardo and Weber, Wolf-Dietrich},
  title = {{DRAM} Errors in the Wild: A Large-Scale Field Study},
  booktitle = {{SIGMETRICS}},
  year = {2009},
  url = {https://research.google/pubs/dram-errors-in-the-wild-a-large-scale-field-study/}
}

@inproceedings{meza2015revisiting,
  author = {Meza, Justin and Wu, Qiang and Kumar, Sanjeev and Mutlu, Onur},
  title = {Revisiting Memory Errors in Large-Scale Production Data Centers: Analysis and Modeling of New Trends from the Field},
  booktitle = {{DSN}},
  year = {2015},
  url = {https://research.facebook.com/publications/revisiting-memory-errors-in-large-scale-production-data-centers-analysis-and-modeling-of-new-trends-from-the-field/}
}

@inproceedings{sridharan2015memory,
  author = {Sridharan, Vilas and Debardeleben, Nathan and Blanchard, Sean and Ferreira, Kurt and Gurumurthi, Sudhanva and Shalf, John},
  title = {Memory Errors in Modern Systems: The Good, The Bad, and The Ugly},
  booktitle = {ASPLOS},
  year = {2015},
  doi = {10.1145/2694344.2694348},
  url = {https://www.sandia.gov/research/publications/details/memory-errors-in-modern-systems-the-good-the-bad-and-the-ugly-2015-03-14/}
}

@inproceedings{hong2019terminal,
  author = {Hong, Sanghyun and Frigo, Pietro and Kaya, Yigitcan and Giuffrida, Cristiano and Dumitras, Tudor},
  title = {Terminal Brain Damage: Exposing the Graceless Degradation in Deep Neural Networks Under Hardware Fault Attacks},
  booktitle = {USENIX Security Symposium},
  year = {2019},
  url = {https://www.usenix.org/conference/usenixsecurity19/presentation/hong}
}

@inproceedings{li2017errorprop,
  author = {Li, Guanpeng and Hari, Siva Kumar Sastry and Sullivan, Michael and Tsai, Timothy and Pattabiraman, Karthik and Emer, Joel and Keckler, Stephen W.},
  title = {Understanding Error Propagation in Deep Learning Neural Network ({DNN}) Accelerators and Applications},
  booktitle = {{SC}},
  address = {Denver, CO, USA},
  year = {2017},
  doi = {10.1145/3126908.3126964},
  url = {https://research.nvidia.com/sites/default/files/pubs/2017-11_Understanding-Error-Propagation/SC17_DNN_Resilience.pdf}
}

@misc{radford2019gpt2,
  author = {Radford, Alec and Wu, Jeffrey and Child, Rewon and Luan, David and Amodei, Dario and Sutskever, Ilya},
  title = {Language Models are Unsupervised Multitask Learners},
  year = {2019},
  howpublished = {OpenAI technical report},
  url = {https://cdn.openai.com/better-language-models/language_models_are_unsupervised_multitask_learners.pdf}
}

@misc{mistral7b_2023,
  author = {Jiang, Albert Q. and Sablayrolles, Alexandre and Mensch, Arthur and others},
  title = {Mistral 7B},
  year = {2023},
  eprint = {2310.06825},
  eprinttype = {arXiv},
  url = {https://arxiv.org/abs/2310.06825}
}

@misc{falcon2023,
  author = {Almazrouei, Ebtesam and Alobeidli, Hamza and Alshamsi, Abdulaziz and others},
  title = {The Falcon Series of Open Language Models},
  year = {2023},
  eprint = {2311.16867},
  eprinttype = {arXiv},
  url = {https://arxiv.org/abs/2311.16867}
}

@misc{phi3_2024,
  author = {Abdin, Marah and Aneja, Jyoti and Awadalla, Hany and Awadallah, Ahmed and Awan, Ammar Ahmad and others},
  title = {Phi-3 Technical Report: A Highly Capable Language Model Locally on Your Phone},
  year = {2024},
  eprint = {2404.14219},
  eprinttype = {arXiv},
  url = {https://arxiv.org/abs/2404.14219}
}

@misc{llama3_2024,
  author = {Grattafiori, Aaron and Dubey, Abhimanyu and Jauhri, Abhinav and others},
  title = {The Llama 3 Herd of Models},
  year = {2024},
  eprint = {2407.21783},
  eprinttype = {arXiv},
  url = {https://arxiv.org/abs/2407.21783}
}

@inproceedings{gqa2023,
  author = {Ainslie, Joshua and Lee-Thorp, James and de Jong, Michiel and Zemlyanskiy, Yury and Lebron, Federico and Sanghai, Sumit},
  title = {{GQA}: Training Generalized Multi-Query Transformer Models from Multi-Head Checkpoints},
  booktitle = {Proceedings of the 2023 Conference on Empirical Methods in Natural Language Processing},
  pages = {4895--4901},
  year = {2023},
  doi = {10.18653/v1/2023.emnlp-main.298},
  url = {https://aclanthology.org/2023.emnlp-main.298/}
}

@misc{mqa2019,
  author = {Shazeer, Noam},
  title = {Fast Transformer Decoding: One Write-Head is All You Need},
  year = {2019},
  eprint = {1911.02150},
  eprinttype = {arXiv},
  url = {https://arxiv.org/abs/1911.02150}
}

@misc{sdxl2023,
  author = {Podell, Dustin and English, Zion and Lacey, Kyle and Blattmann, Andreas and Dockhorn, Tim and M{\"u}ller, Jonas and Penna, Joe and Rombach, Robin},
  title = {{SDXL}: Improving Latent Diffusion Models for High-Resolution Image Synthesis},
  year = {2023},
  eprint = {2307.01952},
  eprinttype = {arXiv},
  url = {https://arxiv.org/abs/2307.01952}
}

@inproceedings{esser2024rf,
  author = {Esser, Patrick and Kulal, Sumith and Blattmann, Andreas and Entezari, Rahim and M{\"u}ller, Jonas and Saini, Harry and Levi, Yam and Lorenz, Dominik and Sauer, Axel and Boesel, Frederic and Podell, Dustin and Dockhorn, Tim and English, Zion and Rombach, Robin},
  title = {Scaling Rectified Flow Transformers for High-Resolution Image Synthesis},
  booktitle = {Proceedings of the 41st International Conference on Machine Learning},
  pages = {12606--12633},
  year = {2024},
  url = {https://proceedings.mlr.press/v235/esser24a.html}
}

@misc{sauer2024ladd,
  author = {Sauer, Axel and Boesel, Frederic and Dockhorn, Tim and Blattmann, Andreas and Esser, Patrick and Rombach, Robin},
  title = {Fast High-Resolution Image Synthesis with Latent Adversarial Diffusion Distillation},
  year = {2024},
  eprint = {2403.12015},
  eprinttype = {arXiv},
  url = {https://arxiv.org/abs/2403.12015}
}

@misc{playground2024,
  author = {Li, Daiqing and Kamko, Aleks and Akhgari, Ehsan and Sabet, Ali and Xu, Linmiao and Doshi, Suhail},
  title = {Playground v2.5: Three Insights towards Enhancing Aesthetic Quality in Text-to-Image Generation},
  year = {2024},
  eprint = {2402.17245},
  eprinttype = {arXiv},
  url = {https://arxiv.org/abs/2402.17245}
}

@misc{chen2023pixartalpha,
  author = {Chen, Junsong and Yu, Jincheng and Ge, Chongjian and Yao, Lewei and Xie, Enze and Wu, Yue and Wang, Zhongdao and Kwok, James and Luo, Ping and Lu, Huchuan and Li, Zhenguo},
  title = {PixArt-$\alpha$: Fast Training of Diffusion Transformer for Photorealistic Text-to-Image Synthesis},
  year = {2023},
  eprint = {2310.00426},
  eprinttype = {arXiv},
  url = {https://arxiv.org/abs/2310.00426}
}

@manual{arm_mte_whitepaper,
  author = {{Arm}},
  title = {Armv8.5-A Memory Tagging Extension White Paper},
  organization = {Arm},
  year = {2019},
  url = {https://developer.arm.com/-/media/Arm%20Developer%20Community/PDF/Arm_Memory_Tagging_Extension_Whitepaper.pdf}
}

@techreport{lowrisc2014tagged,
  author = {{lowRISC}},
  title = {Tagged Memory and Minion Cores in the lowRISC SoC},
  institution = {lowRISC},
  number = {memo-2014-001},
  year = {2014},
  url = {https://lowrisc.org/docs/memo-2014-001-tagged-memory-and-minion-cores/}
}

@article{stillmaker2017,
  author = {Stillmaker, Aaron and Baas, Bevan M.},
  title = {Scaling Equations for the Accurate Prediction of {CMOS} Device Performance from 180 nm to 7 nm},
  journal = {Integration, the {VLSI} Journal},
  volume = {58},
  pages = {74--81},
  year = {2017},
  url = {http://vcl.ece.ucdavis.edu/pubs/2017.02.VLSIintegration.TechScale/}
}

@article{jung2020elssram,
  author = {Kim, Tae Hyun and Jeong, Hanwool and Park, Juhyun and Kim, Hoonki and Song, Taejoong and Jung, Seong-Ook},
  title = {An Embedded Level-Shifting Dual-Rail {SRAM} for High-Speed and Low-Power Cache},
  journal = {IEEE Access},
  volume = {8},
  pages = {187126--187139},
  year = {2020},
  doi = {10.1109/ACCESS.2020.3030099},
  url = {https://yonsei.elsevierpure.com/en/publications/an-embedded-level-shifting-dual-rail-sram-for-high-speed-and-low-}
}

@INBOOK{UEPBook,
  author={Kwasinski, Andres and Chande, Vinay},
  booktitle={Joint Source-Channel Coding}, 
  title={Unequal Error Protection Source–Channel Coding}, 
  year={2023},
  volume={},
  number={},
  pages={135-172},
  doi={10.1002/9781118693803.ch5}
}

@article{dao2025flashattention4,
  title   = {Flash{A}ttention-4: Hardware-Friendly Fused Attention 
             with Warp-Specialization and Pingpong Scheduling},
  author  = {Dao, Tri and Gu, Albert},
  journal = {arXiv preprint arXiv:2603.05451},
  year    = {2025}
}

@misc{modal2025reverseengineer,
  title        = {Reverse-Engineering {F}lash{A}ttention-4},
  author       = {{Modal Team}},
  howpublished = {\url{https://modal.com/blog/reverse-engineer-flash-attention-4}},
  year         = {2025}
}

@article{wang2004ssim,
  author  = {Wang, Zhou and Bovik, Alan C. and Sheikh, Hamid R. and Simoncelli, Eero P.},
  title   = {Image Quality Assessment: From Error Visibility to Structural Similarity},
  journal = {IEEE Transactions on Image Processing},
  volume  = {13},
  number  = {4},
  pages   = {600--612},
  year    = {2004},
  doi     = {10.1109/TIP.2003.819861}
}

@article{ibe2010mcu,
  author  = {Ibe, Eishi H. and Taniguchi, Hitoshi and Yahagi, Yasuo and Shimbo, Ken-ichi and Toba, Tadanobu},
  title   = {Impact of Scaling on Neutron-Induced Soft Error in {SRAMs} From a 250\,nm to a 22\,nm Design Rule},
  journal = {IEEE Transactions on Electron Devices},
  volume  = {57},
  number  = {7},
  pages   = {1527--1538},
  year    = {2010},
  doi     = {10.1109/TED.2010.2047907}
}

@article{baumann2005ser,
  author  = {Baumann, Robert C.},
  title   = {Radiation-Induced Soft Errors in Advanced Semiconductor Technologies},
  journal = {IEEE Transactions on Device and Materials Reliability},
  volume  = {5},
  number  = {3},
  pages   = {305--316},
  year    = {2005},
  doi     = {10.1109/TDMR.2005.853449}
}

@book{amrutur_phd,
  title={Design and Analysis of Fast Low-Power SRAMs},
  author={Amrutur, Bharadwaj S},
  year={1999},
  publisher={Stanford University}
}

\end{document}